\documentclass[a4paper,11pt]{article}

\usepackage{jheppub} 

\usepackage[T1]{fontenc} 
\usepackage{amsmath}
\usepackage{amsthm}
\usepackage{physics}
\usepackage{mathtools}
\usepackage{enumerate}
\usepackage{xcolor}
\usepackage{graphicx}
\usepackage{tikz-feynman}[luatex=false]

\newcommand{\mpl}{M_\mathrm{Pl}}
\newcommand{\VEV}[1]{\left\langle #1 \right\rangle}

\newcommand{\ephi}{\epsilon_{\phi}}
\newcommand{\crit}{\text{crit}}

\allowdisplaybreaks
\title{The Inflated Chern-Simons Number in Spectator Chromo-Natural Inflation}

\author[a]{Hengameh Bagherian,}
\author[a]{Matthew Reece,}
\author[b,c]{and Weishuang  Linda  Xu}

\affiliation[a]{Department of Physics, Harvard University, Cambridge, MA 02138, U.S.A. }

\affiliation[b]{Berkeley Center for Theoretical Physics, University of California, Berkeley, CA 94720, U.S.A.}

\affiliation[c]{Theoretical Physics Group, Lawrence Berkeley National Laboratory, Berkeley, CA 94720, U.S.A.}

\emailAdd{hengameh@g.harvard.edu}

\emailAdd{mreece@g.harvard.edu}

\emailAdd{wlxu@lbl.gov}

\abstract{The chromo-natural inflation (CNI) scenario predicts a potentially detectable chiral gravitational wave signal, generated by a Chern-Simons coupling between a rolling scalar axion field and an SU(2) gauge field with an isotropy-preserving classical background during inflation. However, the generation of this signal requires a very large integer Chern-Simons level, which can be challenging to explain or embed in a UV-complete model. We show that this challenge persists in the phenomenologically viable spectator field CNI (S-CNI) model. Furthermore, we show that a clockwork scenario giving rise to a large integer as a product of small integers can never produce a Chern-Simons level large enough to have successful S-CNI phenomenology. We briefly discuss other constraints on the model, both in effective field theory based on partial-wave unitarity bounds and in quantum gravity based on the Weak Gravity Conjecture, which may be relevant for further explorations of alternative UV completions.
}

\begin{document} 
\maketitle

\section{Introduction}

Inflation is the leading paradigm for the origin of the nearly scale-invariant primordial density perturbations that seeded the formation of structure in the universe. As such, it is potentially a powerful window to physics at energies far above what can be explored in terrestrial experiments. However, we have no clear information on the energy scale at which inflation took place. In conventional single field slow-roll inflation models, a measurement of primordial gravitational waves (tensor modes) would provide such information: both the Hubble expansion rate during inflation and the inflaton field excursion during inflation are proportional to the square root of the tensor-to-scalar ratio $r$~\cite{Lyth:1996im}. A wide range of inflation models share these properties~\cite{Baumann:2011ws, Mirbabayi:2014jqa}. Thus, a near-future measurement of primordial gravitational waves would lead to the conclusion that the inflaton traversed a roughly Planckian range of field values during inflation, and that the Hubble scale during inflation was near $5 \cdot 10^{13}\,\mathrm{GeV}$. It is important to assess the robustness of these conclusions, in order to firmly anchor our knowledge of high-scale inflationary physics relative to the lower energies explored in particle physics.

A leading contender for a qualitatively different source of inflationary gravitational waves arises from models with large, {\em classical} non-abelian gauge field backgrounds during inflation. The original incarnation of such a scenario, ``gauge-flation,'' made the key observation that a classical background for an SU(2) gauge field could spontaneously break the product of (internal) gauge rotations and spatial rotations to the diagonal, preserving isotropy~\cite{Maleknejad:2011jw, Maleknejad:2011sq}. This was shortly followed by a variant model, chromo-natural inflation (CNI), which replaced dimension-eight gauge field interactions in gauge-flation with a dimension-five coupling of an axion $\chi$ to SU(2) gauge fields~\cite{chromo_original} (see~\cite{Adshead:2012qe, Maleknejad:2012dt} for the relationship between the CNI model and the original gauge-flation). Ordinarily, tensor perturbations sourced by gauge field perturbations would be {\em quadratic}, $h_{ij} \sim \delta A_i \cdot \delta A_j$, and hence highly suppressed. The classical gauge field backgrounds in these models, however, allow gauge perturbations to source tensor perturbations {\em linearly}: $h_{ij} \sim A^\mathrm{cl}_{(i} \cdot  \delta A_{j)}$. Furthermore, because the axion coupling to gauge fields leads to tachyonic amplification of one helicity of the gauge field~\cite{Turner:1987bw, Garretson:1992vt}, this interaction sources {\em chiral} gravitational waves~\cite{Adshead:2013qp}, a possibility that received intense attention (e.g.,~\cite{perturbations_adshead,Maleknejad:2014wsa, Numerics_Obata:2016tmo, Maleknejad:whittaker, Obata:2016xcr, Dimastrogiovanni:2016fuu}). At first glance, then, the CNI model seems very appealing: it relies on ingredients (axions coupled to gauge fields through a Chern-Simons interaction) that are ubiquitous in UV completions~\cite{EM_Eta_Adshead}; it provides a novel physical mechanism for sourcing gravitational waves, deviating from the standard logic relating a tensor signal to the scale of inflation; and, fortunately, it can also be distinguished observationally through the intrinsic chirality of its tensor modes. Furthermore, the isotropic SU(2) gauge field background was shown to be an attractor solution, at least beginning from certain anisotropic classical backgrounds~\cite{Maleknejad:2013npa}.

Subsequent investigation of chromo-natural inflation revealed three potentially problematic aspects of the model: inconsistency of the minimal model with observations; backreaction from large perturbations on the dynamics; and a required axion coupling to gauge fields that is large enough to pose a difficulty for UV completions. Our focus in this paper is the third aspect, though to assess its importance we must take into account the others as well. We will now briefly summarize the first two aspects, before explaining the third in more detail.

The first challenge to the original CNI model was a detailed observational one: the parameter space could not simultaneously accommodate the measured value of the spectral index $n_s$ and the upper bound on $r$~\cite{Dimastrogiovanni_Num, Adshead:2013qp,perturbations_adshead}. This spurred the development of modifications to the underlying model, e.g., higgsed chromo-natural inflation, in which the SU(2) gauge fields acquire mass through the Higgs mechanism~\cite{Adshead:2016omu}. Another, more well-studied, modification is the spectator chromo-natural inflation (S-CNI) model, introduced in~\cite{Dimastrogiovanni:2016fuu}. This scenario, which will be our focus in this paper, assumes the existence of a scalar inflaton field in addition to the axion and SU(2) gauge fields of the original CNI model. By decoupling the scale of the inflationary potential from the axion dynamics, this opens up a larger parameter space and relaxes the observational constraints. The S-CNI model was argued to allow for a large gravitational wave signal even for low-scale inflation~\cite{Fujita_2018}. It was also argued to be an attractor solution beginning from anisotropic initial conditions~\cite{Wolfson:2020fqz, Wolfson:2021fya}.

A further challenge to the CNI model arises from backreaction of perturbations. The large gravitational wave signals of interest involve very large occupation numbers of gauge field modes, which source tensor perturbations. The gauge modes can backreact on the classical evolution of the fields, at some point invalidating the perturbation theory, a constraint that removes portions of the parameter space where the sourced contribution to $r$ is much larger than the vacuum contribution to $r$~\cite{Maleknejad_2019}. Furthermore, the tensor perturbations can also source {\em scalar} perturbations via a one-loop diagram, first studied in the original CNI model in~\cite{Papageorgiou:2018rfx} and later in the S-CNI model in~\cite{Papageorgiou:2019ecb}. Accounting for this effect is important to accurately estimate the tensor-to-scalar ratio. Working in a reliably calculable regime with small backreaction, these studies (contrary to the initial ones like~\cite{Fujita_2018}) find severe limitations on the prospects for obtaining a dominantly chiral gravitational wave signal orders of magnitude larger than what one would expect in single field slow-roll inflation with similar Hubble scale. On the other hand, they leave open the possibility of a tensor signal that has an observable chirality, which would be a dramatic signal of dynamics beyond standard single field slow-roll inflation.

In this paper, we assess whether the surviving S-CNI parameter space is feasible from the viewpoint of quantum field theory. Although the basic ingredients of the CNI model are appealingly simple and expected to arise in many UV completions, the detailed parameter choices are more problematic.
A key feature of the CNI model, in common with other models like~\cite{Anber:2009ua}, is the need for a large coupling of axions to gauge fields, of the form $\frac{\lambda}{f} \chi F \widetilde{F}$ where $\chi$ is an axion with a potential of period $2\pi f$ and $\lambda \sim 100$ for phenomenological consistency of the model. As originally pointed out in~\cite{Heidenreich:2017sim}, and elaborated on in~\cite{Reece:Chrono}, this poses a puzzle. An axion is not just any scalar field: it is a periodic scalar, and this implies that such a coupling is in fact a Chern-Simons term with coefficient an {\em integer} multiple of $g^2/(32\pi^2)$. We review this below in Section~\ref{subsec:period}. When $g$ is perturbatively small (as it must be, for consistency of the  model), large $\lambda$ can require that the integer coefficient be enormously large. This requires explanation. In this paper, we will see that this issue is also present in the Spectator CNI model, and again, the large integer cannot be explained with the clockwork mechanism~\cite{Choi:2014rja,Choi:2015fiu,Kaplan:2015fuy}.

In Section~\ref{sec:background} we will review the basic setup of the S-CNI model and establish our notation. In Section~\ref{sec:previousconstraints}, we discuss a number of constraints on the model, arising from compatibility with observations and the existence and control of a slow-roll solution within the effective theory. These constraints have been derived in earlier literature (e.g.,~\cite{Dimastrogiovanni:2016fuu,Maleknejad_2019,Papageorgiou:2019ecb}), although in some cases our treatment is slightly different. After presenting these constraints, we summarize the relatively small viable parameter space that remains. In Section~\ref{section:Lambda}, we provide some useful intermediate technical results to better understand the parameter space and facilitate the subsequent discussion. In Section~\ref{section:clockwork}, we show that the parameters of the S-CNI model cannot be explained with the clockwork mechanism. Section~\ref{section:additional} discusses other constraints on the model, both EFT constraints from perturbative unitarity and (conjectural) UV constraints from embedding in quantum gravity. In Section~\ref{sec:conclusion}, we offer concluding remarks.

\section{The Spectator Chromo-Natural Inflation (S-CNI) Model}
\label{sec:background}

In this section, we review the basic structure of the spectator chromo-natural inflation (S-CNI) model. 

\subsection{S-CNI Ingredients}
    
Let us first review the ingredients of the S-CNI model~\cite{Dimastrogiovanni:2016fuu}. First is the inflaton $\phi$, with potential $V(\phi)$ that is assumed to dominate the energy density of the universe. In particular, $\phi$ should slowly roll for at least 50 to 60 e-folds, and we expect inflation to end when $\phi$ transitions away from slowly rolling, just as in conventional inflation models. Second, we have the chromo-natural sector~\cite{chromo_original}, which consists of an axion field $\chi$ interacting with SU(2) gauge fields $A^a_\mu$ via a Chern-Simons term.\footnote{For the purposes of the model, we could equally well (up to a relative factor of 2 in the allowed normalization of the Chern-Simons term) assume the gauge group to be SO(3).} This sector is assumed to interact with $\phi$ only through gravity. During the observable era of inflation (i.e., those e-folds that sourced the modes that we measure in the CMB), the field $\chi$ is assumed to be rolling in a potential $U(\chi)$ while the gauge field has a classical background that preserves isotropy~\cite{Maleknejad:2011jw}, which we parametrize through a function of time, $Q(t)$: 
\begin{align} \label{eq:Aansatz}
    \expectationvalue{A_0^a(t)} = 0, \qquad \expectationvalue{A_i^a(t)} = \delta^a_i \ a(t) \ Q(t).
\end{align}
Here $a(t)$ is the scale factor, $a$ is an SU(2) adjoint index, and $i$ is a spatial Lorentz index. As emphasized in the S-CNI context in~\cite{Dimastrogiovanni:2016fuu, Papageorgiou:2019ecb}, following similar work in other inflation scenarios~\cite{Barnaby:2012xt, Namba:2015gja}, $\chi$ need not roll for the full duration of $\phi$-driven inflation. The generation of an interesting gravitational wave signal could occur during a limited interval when $\chi$ is rolling. We denote the number of e-folds of axion rolling by $N_\chi$, which we take to be $N_\chi \gtrsim 10$~\cite{Papageorgiou:2019ecb} in order to cover the observationally relevant scales, and assume this epoch was followed by roughly $60 - N_\chi$ additional e-folds of $\phi$-driven inflation. We will not assume any complete model to explain what triggers the axion's rolling at a particular time, but simply focus on the observable signals generated from an EFT capturing the interval in which $\phi$ and $\chi$ both roll. As we will see, this is already highly constrained.

The EFT describing the evolution of $\phi$, $\chi$, and $A^a_\mu$ is assumed to be~\cite{Dimastrogiovanni:2016fuu}
\begin{equation}\label{lagrangian}
    \frac{1}{\sqrt{|\det \mathrm{g}|}} \mathcal{L}=  -\frac{1}{2}(\partial_{\mu}\phi)^2 -V(\phi) 
    -\frac{1}{2}(\partial_{\mu}\chi)^2 - U(\chi) -\frac{1}{4}F_{\mu \nu}^a F^{a\mu\nu}-\frac{\lambda}{4f}\chi F^a_{\mu\nu}\widetilde{F}^{a\mu\nu},
\end{equation}
where $\widetilde{F}^{a \mu \nu} \equiv \frac{1}{2\sqrt{|\det \mathrm{g}|}} \epsilon^{\mu\nu\rho\sigma} F_{\rho \sigma}^a$ and $a \in \{1,2,3\}$ is the SU(2) adjoint index. We denote the SU(2) gauge coupling as $g$, not to be confused with the metric $\mathrm{g}_{\mu \nu}$. We work in a mostly-plus metric signature, and assume a homogeneous, isotropic, background described by the Friedmann-Robertson-Walker (FRW) prescription.  Throughout this paper, we will assume a simple periodic axion potential,
\begin{equation}
    U(\chi) = \mu^4 \left[1 + \cos(\chi/f)\right].
\end{equation}
The scale $\mu$ could arise from confinement in an additional non-abelian gauge sector coupled to $\chi$.     
Because the axion periodicity will play a key role in our discussion below, we will shortly comment on it in more detail, in Section~\ref{subsec:period}.

We adopt the following notation, widely used in the literature, for frequently occurring dimensionless combinations of parameters and field values:
\begin{align} \label{eq:notation}
    m_Q \equiv \frac{g Q}{H}, \qquad \xi \equiv \frac{\lambda}{2fH}\dot \chi, \qquad \Lambda \equiv \frac{\lambda Q}{f}.
\end{align}
Note that $m_Q$ can be thought of as the mass of the gauge field in Hubble units. These are all time-dependent quantities, but can be treated as approximately constant due to slow-roll conditions (to be discussed below in Section~\ref{subsec:eom}).

\subsection{S-CNI Tensor Perturbations}
In order to make predictions regarding the tensor to scalar ratio, we will also need to define the tensor perturbation modes. We are following the notation from Ref.~\cite{Papageorgiou:2019ecb}, to write down the gauge field and metric perturbations:
\begin{align}
    \delta A_{\mu}^1 &= a(x) [0,\ T_+(x, z), \ T_{\cross}(x,z),0 ], &\qquad \delta A_{\mu}^2 & =a(x) [0, \ T_{\cross}(x, z), \ -T_+(x, z), 0 ],\\
    \delta {\rm g}_{12} &= -\delta {\rm g}_{22}=a^2(x) h_+(x, z)/\sqrt{2}, &\qquad \delta {\rm g}_{12}&=\delta {\rm g}_{21}=a^2(x)h_{\cross}(x, z)/\sqrt{2},
\end{align}
where modes are propagating along $z$ and $x\equiv ka/H$. We can now define the left-handed and right-handed helicity tensors as
\begin{align}\label{eq:tensormodes}
    \hat{h}^{\pm} \equiv a\mpl/2\ \left[h_+\mp i h_{\cross}\right], \qquad \hat{t}^{\pm} \equiv a(x)\ \left[T_+ \mp i T_{\cross}\right].
\end{align}
        
\subsection{Axion  periodicity}
\label{subsec:period}

We assume that $\chi$ is a periodic scalar field, i.e., that there is an identification
\begin{equation} \label{eq:axionperiod}
    \chi \cong \chi + 2\pi n f_\chi, \quad n \in \mathbb{Z},
\end{equation}
for some constant $f_\chi$. This is a {\em gauge redundancy}, i.e., these different values of $\chi$ are different ways of referring to literally the same physical field configuration. This is a common feature of (pseudo-)Goldstone bosons, including axions. The periodicity of $\chi$ has important implications for both the potential $U(\chi)$ and the Chern-Simons term $\chi F \widetilde{F}$, previously discussed in the context of the CNI model in~\cite{Reece:Chrono}.

The Chern-Simons term $\chi F \widetilde{F}$ is not gauge-invariant: shifting $\chi$ by $2\pi f_\chi$ adds an effective $\theta$-term to the SU(2) gauge field action. However, all physical quantities are invariant if this added term is an integer multiple of $\frac{g^2}{16\pi} F^a_{\mu \nu} \widetilde{F}^{a\mu\nu}$. The reason is that the integral of this term over spacetime, for any gauge field configuration, is itself $2\pi$ times an integer (the ``instanton number''), so the path integral weight $\exp(i S[\phi,\chi,A])$ is invariant even though the action itself is not. As a result, the coefficient $\lambda/f$ is actually {\em quantized}, i.e., it must be an integer multiple of a base unit:
\begin{equation} \label{eq:CSterminteger}
    \frac{\lambda}{4f} = k \frac{g^2}{32\pi^2} \frac{1}{f_\chi}, \quad k \in \mathbb{Z}.
\end{equation}
The integer $k$ is referred to as the {\em level} of the Chern-Simons term. 

At this stage it is tempting to identify $f_\chi$ with $f$, but we should not do so: there is another subtlety to confront first. If we demand that $U(\chi)$ respects the periodicity~\eqref{eq:axionperiod}, we would conclude that $1/f$ must be an integer multiple of $1/f_\chi$. However, there is a well-known loophole in this argument. In fact, we will allow it to be an {\em inverse} integer multiple:
\begin{equation} \label{eq:Uchiinteger}
    \frac{1}{f} = \frac{1}{j} \frac{1}{f_\chi}, \quad j \in \mathbb{Z}.
\end{equation}
As written, this appears to be a violation of gauge invariance, but in a full model it can be realized as a form of higgsing (spontaneously breaking the gauge invariance). There are many models that exhibit a {\em monodromy}, where as the axion field winds around its circle, another parameter of the theory (e.g., some flux) changes by an integer value. These effects can change the {\em effective} periodicity of the axion potential from $2\pi f_\chi$ to $2\pi j f_\chi$ for an integer $j$, allowing for $j$ different ``branches'' of the potential to restore the true $2\pi f_\chi$ period~\cite{Witten:1980sp, Kim:2004rp}.\footnote{In fact, this is essentially the same mechanism that allows an effectively fractional Chern-Simons level to appear in the fractional quantum Hall effect.}

Putting together~\eqref{eq:CSterminteger} and~\eqref{eq:Uchiinteger}, we learn that
\begin{equation} \label{eq:lambdanorm}
    \lambda = j k \frac{g^2}{8\pi^2}, \quad j, k \in \mathbb{Z}.
\end{equation}
This suggests that the natural value of $\lambda$ is very small, at weak gauge coupling $g$. Only models with a large integer $jk$ can produce a large $\lambda$. This is a fundamental challenge for building a complete CNI or S-CNI model~\cite{Heidenreich:2017sim, Reece:Chrono}.

Here we have assumed that the potential remains periodic, with a finite number $j$ of branches. A qualitatively different scenario can arise in which there are {\em infinitely} many branches, leading to an effectively non-periodic axion potential~\cite{Silverstein:2008sg,Kaloper:2008fb}. We do not consider this case here, sticking with the cosine potential studied in most of the CNI and S-CNI literature. However, more general monodromy potentials have been invoked in similar contexts~\cite{Maleknejad:2016dci, Maleknejad:2020yys, Maleknejad:2020pec}, and could be an interesting target for further study.

One might also ask: why assume that $\chi$ is periodic at all? Perhaps it is a generic scalar with a coupling to $F^a\widetilde{F}^a$, free from the constraints imposed by periodicity. However, in this case, without the protection afforded $\chi$ by being a pseudo-Nambu-Goldstone boson, it would be difficult to explain the flatness of its potential. Furthermore, we are aware of only two basic mechanisms for generating $\chi F^a \widetilde{F}^a$ couplings: one is to integrate out SU(2)-charged fermions whose mass matrix depends on the value of $\chi$, which would lead to similar challenges whether or not $\chi$ is periodic; the second is to obtain $\chi$ as a zero mode of a higher-dimensional gauge field participating in a Chern-Simons interaction with the SU(2) gauge fields, in which case we expect $\chi$ to automatically be periodic. Thus, we do not expect abandoning the periodicity assumption to make the task of UV completing the model any easier.

\subsection{Equations of motion}
\label{subsec:eom}

We can define a set of slow-roll parameters in the S-CNI model, all of which should be much smaller than one during inflation:
\begin{align} \label{eq:slowroll}
    \epsilon_H \equiv -\frac{\dot H}{H^2}, \quad \ephi \equiv \frac{\dot \phi^2}{2\mpl^2 H^2}, \quad \epsilon_{\chi} \equiv \frac{\dot \chi^2}{2\mpl^2 H^2}, \quad \epsilon_B \equiv \frac{g^2Q^4}{\mpl^2 H^2}, \quad \epsilon_E \equiv \frac{\big(HQ+\dot Q\big)^2}{H^2\mpl^2}.
\end{align}
Here we use a dot to denote a derivative with respect to the FRW time coordinate $t$, $H$ is the Hubble parameter $H(t) \equiv {\dot a}(t)/a(t)$, and $\mpl \equiv \frac{1}{\sqrt{8\pi G_N}} \approx 2.4 \cdot 10^{18}\,\mathrm{GeV}$ is the reduced Planck scale.
We also define $\epsilon_A \equiv \epsilon_B + \epsilon_E$.
There are also slow-roll conditions on second derivatives:
\begin{align} \label{eq:secondderivs}
    |\ddot \phi| \ll |H\dot \phi|, \quad |\ddot \chi| \ll |H\dot \chi|, \quad |\ddot Q| \ll |H\dot Q|.
\end{align}
    
Before using any approximations, the classical background equations of motion for the fields $\phi$, $\chi$, and $Q$ are
\begin{align}
    \ddot{\phi}& + 3 H \dot\phi + V'(\phi) = 0,\label{eq:phieomorig}\\
    \ddot{\chi}&+3H\dot\chi+U'(\chi)=-\frac{3g\lambda}{f}Q^2\left(\dot{Q}+HQ\right),\label{xieom}\\
    \ddot{Q}&+3H\left(\dot{Q}+HQ\right)-H^2Q\left(1-\dot H/H^2\right)+2g^2Q^3 = \frac{g\lambda}{f}\dot\chi Q^2 \label{Qeom}.
\end{align}
In these equations, a prime denotes a derivative of a function with respect to its argument.
These must be supplemented with one linear combination of the Einstein equations; a useful choice can be expressed in terms of the slow-roll parameters~\eqref{eq:slowroll}:
\begin{equation} \label{eq:slowrollsum}
    \epsilon_H = \epsilon_\chi + \epsilon_\phi + \epsilon_A,
\end{equation}
 which also allows for straightforward assessment of the relative importance of the inflaton, axion, and gauge field contributions on the dynamics of inflation. The four equations~\eqref{eq:phieomorig},~\eqref{xieom},~\eqref{Qeom}, and~\eqref{eq:slowrollsum} fully determine the evolution of the classical background, assuming an FRW metric and the isotropic SU(2) ansatz~\eqref{eq:Aansatz}.

These equations depend on a choice of the inflaton potential $V(\phi)$. However, we will follow~\cite{Dimastrogiovanni:2016fuu} in not choosing an explicit inflaton potential $V(\phi)$. Instead, we will replace the equation of motion~\eqref{eq:phieomorig} with an approximate equation of motion for the slow-roll parameter $\ephi$. The dynamics of $\ephi$ are fixed by observed properties of the primordial density fluctuations, which we assume are sourced dominantly by $\phi$, allowing us to remain agnostic to the underlying potential. The amplitude of the scalar power spectrum can be written in terms of $\ephi$ and $\epsilon_H$, derived in~\cite{Papageorgiou:2019ecb} as 
    \begin{equation} \label{eq:scalarpower}
        \mathcal{P}_{\zeta} = \frac{H^2}{8\pi^2\mpl^2} \frac{\ephi}{\epsilon_H^2},
    \end{equation}
while the time-evolution of $\ephi$ may be inferred from the spectral tilt of the spectrum, taken as approximately constant, $n_s \simeq 0.965 \pm 0.004$~\cite{Planck:2018vyg}. This is given explicitly by 
\begin{align}\label{eq:phieom}
    \dot{\epsilon}_{\phi}&\approx H\ephi\Big[-6\dot H/H^2+\left(n_s-1\right)+2\ddot H/(\dot H H) \Big],
\end{align}
and we refer the reader to  Appendix~\ref{app:phieom} for details of this derivation. Note that underlying this entire prescription is the assumption that slow-roll conditions are met.
We treat~\eqref{xieom},~\eqref{Qeom},~\eqref{eq:slowrollsum}, and~\eqref{eq:phieom} (rather than~\eqref{eq:phieomorig}) as the four equations specifying the background evolution.

After using the slow-roll assumptions to drop the second derivatives, the two equations~\eqref{xieom} and~\eqref{Qeom} can be separated into an equation for $\dot \chi$ and one for $\dot Q$. It is useful to think of the equation of motion for $\dot Q$ as an equation for gradient flow in an effective potential $W_\mathrm{eff}(Q)$ (with $\chi$ and $H$ treated, for this purpose, as approximately constant)~\cite{chromo_original,Dimastrogiovanni_Num}. Specifically,
    \begin{equation} \label{eq:Qdot}
        {\dot Q} \approx -\frac{\partial W_\mathrm{eff}(Q)}{\partial Q} = -\frac{3HQ (2 f^2 H^2 + 2 g^2 f^2 Q^2 + g^2 \lambda^2 Q^4) + g \lambda Q^2 f U'(\chi)}{3(3 f^2 H^2 + g^2 \lambda^2 Q^4)}.
    \end{equation}
The solutions of interest for us have $Q$ approximately sitting at a nonzero minimum of this effective potential.

The $\chi$ equation of motion leads to an important conclusion about the necessary size of the parameter $\lambda$. Using the slow-roll equation for $\dot \chi$ and rewriting $U'(\chi)$ in terms of other parameters by assuming that $Q$ sits at a value where the right-hand side of~\eqref{eq:Qdot} is zero, we find:
\begin{equation} \label{eq:chidot}
    {\dot \chi} \approx \frac{2 H f}{\lambda} \left(m_Q + m_Q^{-1}\right).
\end{equation}
From the definition of $\xi$ in~\eqref{eq:notation}, this is equivalent to the claim
\begin{equation} \label{eq:xiestimate}
    \xi \approx m_Q + m_Q^{-1}. 
\end{equation}
The expression~\eqref{eq:chidot} makes it clear that large $\lambda$ suppresses the evolution of $\dot \chi$; this is the key friction effect that causes the time evolution of the axion to differ in the presence of the SU(2) gauge fields compared to an ordinary axion, allowing the CNI model to have sub-Planckian axion inflation (in contrast to natural inflation~\cite{Freese:1990rb}). On the other hand, we also see from~\eqref{eq:chidot} that in the {\em small} $\lambda$ limit, $\dot \chi$ becomes large: the axion evolves even faster than it otherwise would.

Ignoring the time dependence of $Q$, we can roughly estimate the number of e-folds of CNI dynamics from~\eqref{eq:chidot} simply by asking how many Hubble times it takes for $\chi$ to change value by $f$. We see that
\begin{equation}
    N_\chi \sim \frac{\lambda}{2 \left(m_Q + m_Q^{-1}\right)}.
\end{equation}
This derivation is valid in the S-CNI model, but leads to similar conclusions as in the original CNI model (e.g.,~\cite{chromo_original} states that $N_\chi \lesssim 0.6\lambda$). Requiring a number of e-folds $N_\chi \gtrsim 10$ translates into the demand that
\begin{equation} \label{eq:lambdarequirement}
    \lambda \sim 2 \left(m_Q + m_Q^{-1}\right) N_\chi \gtrsim 60,
\end{equation}
where we have used that (as explained below) $m_Q \sim 2.7$ in the viable parameter space. Thus, we are only interested in parameter space with large coupling $\lambda$. In terms of the parametrization~\eqref{eq:lambdanorm}, this means that
\begin{equation}
    j k = \frac{8\pi^2 \lambda}{g^2} \gtrsim 5 \cdot 10^7 \left(\frac{10^{-2}}{g}\right)^2.
\end{equation}
This large integer proves to be a significant challenge for UV completions of the model, as we will see below.

\section{Constraints from the EFT}
\label{sec:previousconstraints}

As explored in previous work~\cite{Adshead:2012qe,Maleknejad_2019, Dimastrogiovanni:2016fuu, Papageorgiou:2019ecb}, multiple constraints and inequalities split and restrain the available space of parameters for the S-CNI model. In this section we will discuss and expand on some of these bounds. We put these constraints in two broad categories: compatibility with observational bounds and physical control over the behavior of solutions. 

The set of key parameters that we will mainly use to dictate the observable behavior of S-CNI are $H$, $g$, and $m_Q$. These govern the energy content stored in the gauge field and (via subtraction from Hubble) in the inflaton. The prominence of the axion as a player in inflationary phenomenology, however, is governed by an additional parameter $\Lambda$ (defined in~\eqref{eq:notation}), which encodes dependence on the axion-related parameters: $\lambda$, $f$, and (by way of equations of motion) $\mu$. For sufficiently small $\Lambda$, the axion becomes dynamic enough to impact the evolution of density perturbations. 

This parameter $\Lambda$ has been traditionally taken to its large limit $\Lambda \gg \sqrt{2}$ and decoupled from the analysis, as 
analytically solving the equations of motion for $Q$ significantly simplifies in this limit. In addition, in the original CNI model having a large $\Lambda$ was necessary to drive a sufficiently long phase of slow-roll inflation \cite{Dimastrogiovanni_Num,Dimastrogiovanni:2016fuu, Papageorgiou:2019ecb}. Thanks to the addition of the new spectator scalar, however, S-CNI doesn't require large values of $\Lambda$ to last for enough e-folds. This opens up an unexplored region for CNI dynamics. However, because of the aforementioned analytic appeal of the $\sqrt{2} \ll \Lambda$ regime and the similarities in the behavior of the axion--gauge sector to that of the original model, most of the S-CNI literature has been focused on large $\Lambda$ regimes. 

In this work, we analyze the full range of values for $\Lambda$, and in later sections explore the limitations of the clockwork mechanism in this space. Hence, it is important to keep track of the $\Lambda$ dependence of each of our bounds as we come across them in this section.

\subsection{Agreement with Observational Data}

In this section, we discuss constraints that need to be met if the model is to comply with experimental data. There are, generically speaking, three cosmological observables that inflation-era physics must confer with: the ratio between primordial tensor and scalar fluctuations $r$, the amplitude of adiabatic scalar modes $\mathcal{P}_\zeta$, and its spectral tilt $n_s$.  We use the measured value of $n_s$ to fix the evolution of the spectator inflaton $\phi$ (see Appendix~\ref{app:phieom} for details), and so a spectral index  consistent with the data is a presupposed part of our S-CNI phenomenology. In the text following, we discuss constraints from the other two measurements.

Recent data from the BICEP/Keck collaboration have placed stringent limits on the tensor-to-scalar ratio, $r < 0.036$ \cite{BICEP:2021xfz} at the 95\% CL.  Here, we summarize the derivation of $r$ under the S-CNI framework, following closely the work of Ref.~\cite{Dimastrogiovanni:2016fuu}. 

The metric perturbation modes $\hat{h}^{\pm}$ are sourced in part by the tensor perturbations of the gauge field $\hat{t}^{\pm}$, both defined in Sec.~\ref{sec:background}. An instability of the $+$-helicity gauge field tensor mode during horizon crossing induces a large enhancement in the metric fluctuations of the same helicity. As a result, the observable tensor modes are likewise highly enhanced, and the enhanced component is entirely chiral.

The power spectrum of these sourced gravitational modes at the super-horizon limit is 
\begin{align}\label{eq:GWpower_def}
    \mathcal{P}_h^s = \frac{\mathcal{P}_h^v}{2}\  \big| \sqrt{2k}x \lim_{x\to 0} \hat{h}^{+}(x,k) \big|^2,
\end{align}
where $\mathcal{P}_h^v = \frac{2}{\pi^2}\frac{H^2}{\mpl^2}$ is the amplitude of the gravitational power spectrum produced through vacuum fluctuation; the measurement of $r$ probes the sum total amplitude of both the vacuum and sourced contributions. 

The tensor fluctuations of the gauge field evolve as $\hat{t}^{+} = i^{\beta}(2k)^{-1/2} W_{\beta, \alpha}(-2ix)$. The $W_{\beta,\alpha}(z)$ are Whittaker functions with arguments $\alpha \equiv -i\sqrt{2m_Q\xi-1/4}$ and $\beta \equiv -i(m_Q+\xi)$. Using Bunch-Davies vacuum solutions to construct the Green's function for $\hat{h}^{+}$ and integrating it over the Whittaker source terms (see Appendix E of~\cite{Maleknejad_2019} and Appendix B of~\cite{Maleknejad:whittaker}), we arrive at an analytic expression
\begin{equation}\label{eq:F2}
    \big| \sqrt{2k}x \lim_{x\to 0} \hat{h}^{+}(x,k) \big|^2 = |\mathcal{F}_E \sqrt{\epsilon_E}+\mathcal{F}_B\sqrt{\epsilon_B}|^2 \equiv \epsilon_B\mathcal{F}^2,
\end{equation}
with 
\begin{align}
    \mathcal{F}_E = &\frac{\pi  i^{\beta +1} \sec (\pi  \alpha )\left[16 \alpha ^4-40 \alpha ^2+9\right]^{-1}}{ \Gamma (1-\beta ) \Gamma \left(-\alpha -\beta +\frac{1}{2}\right) \Gamma \left(\alpha -\beta +\frac{1}{2}\right)} \bigg[16 \left(4 \alpha ^2-8 \beta -1\right) \Gamma (1-\beta )^2\\\nonumber
    &+\left[-16 \alpha ^4+8 \alpha ^2 (5-8 \beta )+16 \beta  (1-8 \beta )-9\right] \Gamma \left(-\alpha -\beta +1/2\right) \Gamma \left(\alpha -\beta +1/2\right)\\\nonumber
    &+\left(16 \alpha ^4-40 \alpha ^2+9\right) \Gamma (-\beta ) \Gamma (1-\beta ) \bigg],
\end{align}
and 
\begin{align}\label{eq:FB}
    \mathcal{F}_B = \frac{\pi  i^{\beta } \sec (\pi  \alpha )\left[-4 \alpha ^2-8 i \beta  m_Q+9\right] \left[\frac{4 \alpha ^2+8 \beta -1}{\Gamma (1-\beta )}+\frac{\left(4 \alpha ^2-8 \beta -1\right) \Gamma (-\beta )}{\Gamma \left(-\alpha -\beta +\frac{1}{2}\right) \Gamma \left(\alpha -\beta +\frac{1}{2}\right)}\right]}{16 \alpha ^4-40 \alpha ^2+9}.
\end{align}

We refer the reader to~\cite{Dimastrogiovanni:2016fuu} for further details, though note that our analytical expression $\mathcal{F}_B$ is slightly different due to an assumed typo in their text\footnote{Our result closely follows their numerical fit  $\mathcal{F}^2 \sim \text{exp}(3.6 m_Q)$ (also in~\cite{Papageorgiou:2019ecb}) and reproduces their Figures, so the disagreement must be a spurious one.}.

These expressions, along with Eqs.~\eqref{eq:GWpower_def},~\eqref{eq:F2}, allow us a direct  determination of $r$ for a given $g, m_Q$, and $H$. The bound on tensor-to-scalar ratio thus translates to a bound on this parameter space,
\begin{equation}\label{eq:tensortoscalr}\tag{CON.1}
    r = \frac{\mathcal{P}_h^v+\mathcal{P}_h^s}{\mathcal{P}_{\zeta}} = r_\mathrm{vac}(1+\frac{\epsilon_B}{2}  \mathcal{F}^2) < 0.036.
\end{equation}

As the characteristic signature of CNI scenarios are these chiral tensor modes, model realizations where they constitute a larger fraction of the allowed total are more phenomenologically interesting to pursue. We highlight regions of parameter space where the contributions of the sourced gravitational waves dominates over vacuum contributions, 
\begin{equation}\label{RGG1}\tag{CON.2}
    \mathcal{R}_{GW}=\frac{r_s}{r_\mathrm{vac}}= \frac{\epsilon_B}{2}\mathcal{F}^2 >1.
\end{equation} 
The derivation of~\eqref{eq:tensortoscalr} and~\eqref{RGG1} relied on slow-roll assumptions but not on the large-$\Lambda$ approximation, and are sensitive only to the choice of $g, m_Q$ and $H$ (or equivalently $r_{\rm vac}$).  To see this dependence explicitly, we can use definitions~\eqref{eq:notation} and~\eqref{eq:slowroll} to write ~\eqref{RGG1}  as 
\begin{align}\tag{CON.$2^*$}\label{eq:RGW_Prime}
    \mathcal{R}_{GW} = \frac{\epsilon_B}{2}\mathcal{F}^2 = \frac{m_Q^4}{g^2} \frac{H^2}{\mpl^2} \frac{\mathcal{F}^2}{2}= \frac{m_Q^4}{g^2}\frac{\pi^2}{2}r_{\text{vac}}\mathcal{P}_{\zeta} \ \frac{\mathcal{F}^2}{2} >1,
\end{align}
with $\mathcal{P}_{\zeta}^{\rm obs} \simeq 2.1 \cdot 10^{-9}$~\cite{Planck:2018vyg} being the amplitude of the scalar power spectrum.

To this point, our model must also produce the correct amount of scalar fluctuations  $\mathcal{P}_{\zeta}$. 
The spectator inflation scenario assumes that both the scale of inflation and size of scalar fluctuations are set dominantly by the inflaton $\phi$. However, the evolution of these perturbations are governed also by the {\emph dynamics} of Hubble, $\epsilon_H = -\dot H/H^2$, to which $\ephi$ need not be the dominant contributor.  The expression for the scalar power spectrum was generalized along precisely these lines by Ref.~\cite{Papageorgiou:2019ecb}, giving 
\begin{equation}
    \mathcal{P}_{\zeta} = \frac{H^2}{8\pi^2\mpl^2} \frac{\ephi}{\epsilon_H^2}.
\end{equation}
Substituting $\epsilon_H = \epsilon_{\chi}+\epsilon_A + \ephi$, we obtain a requirement for $\ephi$ that is entirely agnostic of underlying inflaton model, 
\begin{align}\label{eq:branches}\tag{CON.3}
    \ephi = \frac{\epsilon_A + \epsilon_{\chi}}{2}\left[A-2\pm \sqrt{A\ (A-4)} \ \right], \quad \text{with } 4 \leq A \equiv \frac{H^2}{8\pi^2\mpl^2\mathcal{P}_{\zeta}}\frac{1}{\epsilon_A+\epsilon_{\chi}}.
\end{align}
Given a fixed inflation scale $H$ and $\mathcal{P}_{\zeta}$, the analytic constraint $4\leq A$ puts an upper bound on $\epsilon_A + \epsilon_{\chi}$ to ensure the existence of a slow-roll solution to our equations. In other words, ~\eqref{eq:branches} ensures the predicted value for the scalar power spectrum is correctly normalized with respect to the observed $\mathcal{P}_{\zeta}^\mathrm{obs} \simeq 2.1 \cdot 10^{-9}$ for given slow-roll parameters. Intuitively, if the evolution of Hubble is too much faster than the rolling of the inflaton, the inflaton perturbations are too washed-out to constitute our observed primordial inhomogeneities.
    
The two solutions for $\ephi$ correspond to regimes  $\frac{\ephi}{\epsilon_A + \epsilon_{\chi}} >1$ and $\frac{\ephi}{\epsilon_A + \epsilon_{\chi}}<1$, which split the parameter space of the S-CNI model~\cite{Papageorgiou:2019ecb}. In this paper we call the former branch \textit{classic} S-CNI while the latter is named \textit{mixed} S-CNI. In classic S-CNI, the inflaton both dominates the energy density of the universe and is responsible for driving inflation forward. In mixed S-CNI, the inflaton still dominates the energy density, but the dynamics of inflation are dictated by the gauge field. Both branches show attractor solution behavior. 

The $\Lambda$-dependence of~\eqref{eq:branches} plays a significant role in our future discussions. Taking $\dot Q \approx 0$, and using definitions~\eqref{eq:notation} and ~\eqref{eq:slowroll}, we can write
\begin{align}\label{eq:branches_Ldep}
    \epsilon_A + \epsilon_{\chi} = \frac{H^2/\mpl^2}{g^2}\bigg(m_Q^4+m_Q^2+\frac{2(1+m_Q^2)^2}{\Lambda^2}\bigg).
\end{align}
As we discussed in Section~\ref{subsec:eom}, $\chi$ rolls quickly when $\lambda$ is small, as reflected here in the growth of $\epsilon_\chi$ at small $\Lambda \propto \lambda$. In the opposite limit, $\Lambda \gg \sqrt{2}$, $\epsilon_\chi$ is suppressed and setting $\epsilon_H \simeq \ephi+\epsilon_B$ is a good approximation (as one notes that  $\epsilon_E/\epsilon_B \simeq m_Q^{-2}$, and $m_Q > \sqrt{2}$ as discussed below). We can then replace~\eqref{eq:branches} with 
\begin{align}\label{eq:branchesApprox}\tag{CON.$3^*$}
    4 \lesssim \frac{H^2}{8\pi^2\mpl^2\mathcal{P}_{\zeta}}\frac{1}{\epsilon_B} = \frac{g^2}{8\pi^2\mathcal{P}_{\zeta}m_Q^4}
\end{align}
This approximation breaks down when $\epsilon_{\chi}$ can no longer be ignored in the sum $\epsilon_H = \ephi+\epsilon_A + \epsilon_{\chi}$ at $\Lambda \sim \sqrt{2}$.

\subsection{Physical Control}

Our second category is physical consistency;  these constraints restrict us to regimes where our control over the theory is robust and our solutions self-consistent.
        
First and foremost, we assume that inflation is well-described by slow-roll, in the regime where Eqs.~\eqref{eq:slowroll} and~\eqref{eq:secondderivs} are valid.

Next, we avoid tachyonic \emph{scalar} perturbations on sub-horizon scales by putting a lower bound on $m_Q$ (defined in~\eqref{eq:notation})~\cite{Adshead:2013qp}, 
\begin{equation}\label{eq:notachyon}\tag{CON.4}
    \sqrt{2} < m_Q\ .
\end{equation}
At all values of $\Lambda$, this condition is necessary to have a theory that is well-described by an approximate classical solution with a nearly scale-invariant spectrum of density perturbations.
        
A major approximation taken for feasibility of computation is to neglect the backreaction of enhanced tensor perturbations on the background gauge fields. As derived in Refs.~\cite{Dimastrogiovanni:2016fuu,Papageorgiou:2019ecb}, the backreaction term altering the right hand side of Eq.~\eqref{Qeom}, is
\begin{align}\label{backreacterm}
    \mathcal{T}_{BR}^Q = gH^3\xi/(12\pi^2) \ \big[ \mathcal{B}(m_Q) - \tilde{\mathcal{B}}(m_Q)/\xi\big],
\end{align}
where $\xi$ was defined in~\eqref{eq:notation}. In the above equation 
\begin{align}
    \mathcal{B} \equiv \int_{x_{\min}}^{x_{\max}} dx \ x \big | i^{-\alpha}W_{\alpha, \beta}(2ix) \big|^2, \qquad
    \tilde{\mathcal{B}}\equiv \int_{x_{\min}}^{x_{\max}} dx \ x^2 \big|i^{-\alpha} W_{\alpha, \beta}(2ix)\big|^2,
\end{align}
where $x_{\min \text{/} \max} = m_Q + \xi \pm \sqrt{m_Q^2+\xi^2}\ $ is the interval in which the tensor mode undergoes tachyonic enhancement. We will use the numerical fit $\mathcal{B}(m_Q) - \tilde{\mathcal{B}}(m_Q)/\xi \simeq 2.3 \cdot e^{3.9 m_Q}$ to facilitate our computations throughout this paper~\cite{Papageorgiou:2019ecb}. 
        
To quantify the effects of backreaction, we rewrite~\eqref{Qeom} in terms of $\partial V_{\text{eff}}(Q)/\partial Q \equiv 2H^2Q(1+m_Q^2)-\frac{g\lambda}{f}\dot{\chi}Q^2$ \cite{Maleknejad_2019}:
\begin{align}\label{eq:bornapp}
    \ddot Q+ 3H \dot Q + \dot H Q + \partial V_{\text{eff}}/\partial Q \simeq -\mathcal{T}_{BR}^Q.
\end{align}
Under slow-roll assumptions, we ignore $\big|\ddot Q/H\dot Q\big|$ and set $\partial V_{\text{eff}}/\partial Q \simeq 0$. If we further assume $\epsilon_H = -\dot H/ H^2\simeq \text{const}$, Eq.~\eqref{eq:bornapp} can be solved as a standard non-homogeneous first order differential equation. The homogeneous solution to~\eqref{eq:bornapp} is $Q^{(0)}(t) = Q(1-\dot H/3H t)$ which leads to the full solution
\begin{align}
    Q^{(1)}(t) - Q &\simeq -\frac{\mathcal{T}_{BR}^{Q}}{3H}\ t\Big[1+\frac{\dot H}{\mathcal{T}_{BR}^{Q}}Q-\big(2m_{Q}\xi^{-1}+3.9 m_{Q}\big)\frac{\dot H}{3H}t\Big] \simeq  -\frac{\mathcal{T}_{BR}^{Q}}{3H}t,
\end{align}
where $Q$ sets the right hand side of~\eqref{eq:Qdot} to zero. Now, keeping the backreaction small corresponds to
\begin{align}\label{timeclump}\tag{CON.5}
    \big|Q^{(1)}(t)-Q\big|\cdot Q^{-1} \simeq \Big| -\frac{\mathcal{T}_{BR}^{Q}}{3H}\frac{t}{Q} \Big| = \frac{2}{3} N_{\chi} \cdot  \bigg| \frac{g^2}{24 \pi^2} \frac{\mathcal{B}(m_{Q}) - \tilde{\mathcal{B}}(m_{Q})/\xi_*}{m_{Q}/\xi} \bigg| \ll 1,
\end{align}
where $N_{\chi} \sim 10$ is the number of e-folds that the axion rolls. At first glance,~\eqref{timeclump} differs from that of  Ref.~\cite{Papageorgiou:2019ecb} by a factor of $\sim 6$. However, note that their expression is time dependent, whereas~\eqref{timeclump} has been integrated over the period of axion rolling. The extra factor keeps track of backreaction that accumulates over time. In this paper, we will be using this slightly modified backreaction constraint with the extra factor included. Note that no $\Lambda$ assumptions were invoked during this discussion, and so~\eqref{timeclump} holds for the full range of $\Lambda$.
                
                
Our last consistency constraint comes directly from Ref.~\cite{Papageorgiou:2019ecb} and ensures quantum effects can be safely ignored in our classical field backgrounds:
\begin{align}\label{eq:noquantum}\tag{CON.6}
    \mathcal{R}_{\delta \phi}\equiv \frac{\expectationvalue{\delta \phi^{(s)}\delta \phi^{(s)}}}{\expectationvalue{\delta \phi^{(v)}\delta\phi^{(v)}}}\simeq \frac{5\cdot 10^{-12}}{\left( 1+\frac{\epsilon_A}{\ephi}\right)^2} \ e^{7m_Q} \ m_Q^{11}\ N_{\chi}^2\ r_\mathrm{vac}^2<1.
\end{align}

The $\Lambda$ dependence of~\eqref{eq:noquantum} is more subtle. In deriving~\eqref{eq:noquantum}, taking $\Lambda \gg 1$ is a built-in assumption because the non-linear contributions of the axion to the scalar power spectrum only dominate over linear contributions when $\Lambda^2 \gg 1$~\cite{Papageorgiou:2019ecb}.  This leads us to believe the contributions of~\eqref{eq:noquantum} are negligible when $\Lambda < 1$. Incidentally,~\eqref{eq:noquantum} is the only constraint that behaves differently in the classic S-CNI vs.~mixed S-CNI branches, because the expression substituted for $\ephi$ is different for two branches (see~\eqref{eq:branches}). We, therefore, expect the two branches of S-CNI to show similar behaviors in the $\Lambda <\sqrt{2}$ regime where~\eqref{eq:noquantum} is irrelevant. 

\subsection{Low Energy Parameter Space Exploration}\label{sec:lowEFTsumm}

Now, armed with observational bounds and physical  constraints, ensuring well-behaved solutions, we begin to explore the space that meets all of these criteria.

We start analyzing the effects of our combined constraints in the $\Lambda\gg \sqrt{2}$ region first. The approximate constraint given in~\eqref{eq:branchesApprox}, valid in this regime, is only a function of variables $g$ and $m_Q$. We thus obtain the relation $m_Q \leq \big( 32\pi^2\mathcal{P}_{\zeta}\big)^{-1/4} \sqrt{g}\ \simeq \ 35\sqrt{g}$. This, in conjunction with~\eqref{eq:notachyon}, fixes the value of viable $m_Q$ within the $g$-dependent region  $\sqrt{2} < m_Q \lesssim 35 \sqrt{g}$ \cite{Papageorgiou:2019ecb}, as well as providing a lower-bound on the value for $g$, $\left(\frac{\sqrt{2}}{35}\right)^2 \simeq 1.6 \cdot 10^{-3} \lesssim g$. Along the same slice of parameter space we may impose~\eqref{timeclump} to restrict to regions with controlled backreaction.

\begin{figure}[t]
    \centering
    \begin{minipage}{0.5\textwidth}
        \centering
        \includegraphics[width=\textwidth]{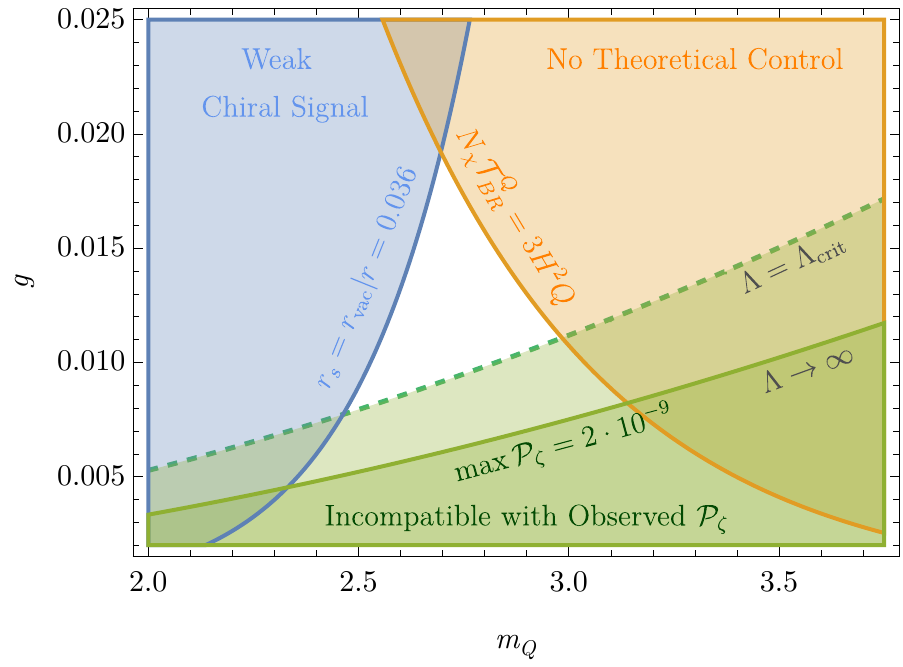}
    \end{minipage}\hfill
    \begin{minipage}{0.49\textwidth}
    \centering
        \includegraphics[width=\textwidth]{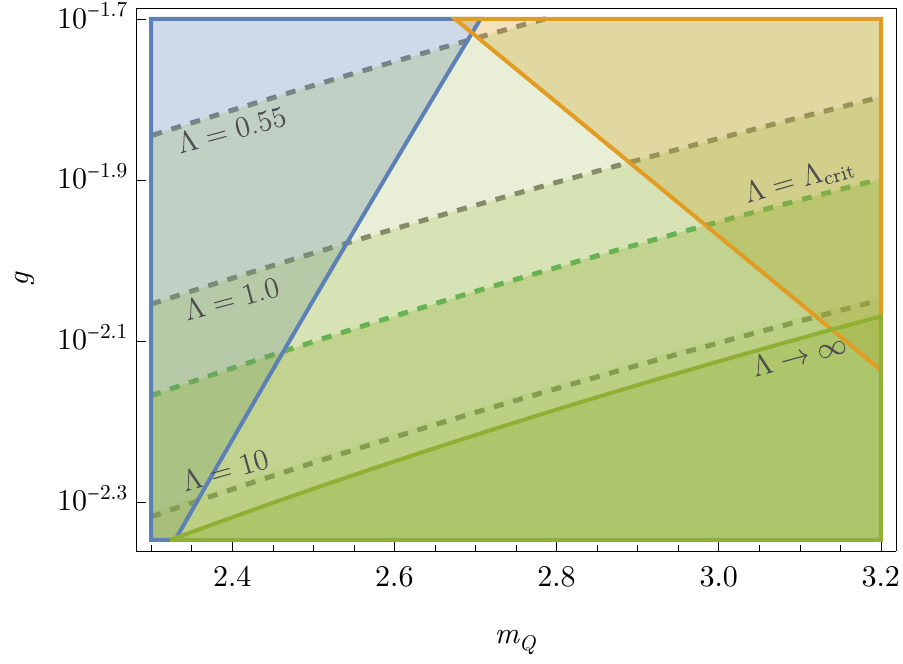}
    \end{minipage}
    \caption{The various constraints from the low energy EFT are superposed to determine the allowed $m_Q-g$ parameter space. \eqref{eq:notachyon}, $\sqrt{2} < m_Q$, is satisfied everywhere on both panels. The blue shaded region doesn't provide a strong chiral gravitational signal and is in violation of~\eqref{eq:RGW_Prime}. The orange shaded region experiences too much contamination from the backreaction~\eqref{timeclump}. The green shaded region with a solid border is incompatible with~\eqref{eq:branchesApprox} and only predicts scalar power spectrum values that are smaller than $\mathcal{P}_{\zeta}^\mathrm{obs} = 2.1 \cdot 10^{-9}$. Its solid green border corresponds to $\Lambda \to \infty$. \textbf{(Left Panel)} Only a small white region doesn't violate any of the constraints. As we lower $\Lambda$ the solid green border moves higher and reduces the available parameter space (see~\eqref{eq:branches_Ldep}). The green shaded region with a dashed border breaks~\eqref{eq:branches} for values of $\Lambda \geq \Lambda_{\crit}\sim \sqrt{2}$ (see~\eqref{eq:Lambda1lowerbound}). \textbf{(Right Panel)} The viable region for $m_Q-g$ values (see left panel) superposed with dashed gray $\Lambda_{\min} = \text{const.}$~contours that saturate~\eqref{eq:Lambdalowerbound}. For any pair $(m_Q, g)$, there is a limit on how low $\Lambda$ can go and still accommodate the correct value for  $\mathcal{P}_{\zeta}$. No choice of $(m_Q, g)$ allows for $\Lambda\lesssim 0.55$, whereas $\Lambda \to \infty$ represents the irreducible, conservative, ruled-out region of parameter space where no value of $\Lambda$ is viable. }\label{fig:gmq} 
\end{figure}

Finally, we consider~\eqref{eq:tensortoscalr} and~\eqref{RGG1} along these axes: for fixed $g$ and $m_Q$ the value of $\epsilon_B$ is determined by the scale of inflation $\epsilon_B \sim H^2$. Thus the sourced contribution to $r$ inherits an even steeper dependence on $H$ than its vacuum counterpart, $r_{\rm vac} \sim H^2$ vs.~$r_{s} \sim H^4$ -- the strength of distinctive S-CNI signatures is enhanced significantly for higher scales of inflation. $H$ is, however, forbidden from becoming arbitrarily large by the observational bound~\eqref{eq:tensortoscalr}, or equivalently the saturation of this bound sets a maximum scale of inflation for each $\{g, m_Q \}$.  This in turn sets a maximally viable signal ratio $\mathcal{R}_{\rm GW}$, to which we impose~\eqref{RGG1} for a conservative constraint on this space. This method of obtaining a maximally viable~\eqref{RGG1} is equivalent to setting $r_{\text{vac}}= 0.018$ in~\eqref{eq:RGW_Prime}. The synthesis of these various constraints, explored in Figure~\ref{fig:gmq} (left panel), delineates a highly confined region of $g$ and $m_Q$ where observable, predictive, and viable S-CNI theories may reside. 

We can read off the small allowed range of $g$ and $m_Q$ as
\begin{align} \label{eq:grange}
    4.5 \cdot 10^{-3} \lesssim &\ g \lesssim 1.9 \cdot 10^{-2}, \\
    \label{eq:mqrange}
    2.3 \lesssim &\ m_Q \lesssim 3.1 \ .
\end{align}
These ranges are similar to those quoted in~\cite{Papageorgiou:2019ecb}, but differ slightly because of the updated observational bound on $r$, our new expression for $\mathcal{F}^2$, and our inclusion of the time integration factor in the backreaction constraint. Intuitively speaking, too large of a coupling generates too much backreaction or too little tensor perturbation. On the other hand, too small of a coupling causes the gauge field to be too light to source $\mathcal{P}_{\zeta}$. 


As we move away from the large $\Lambda$ limit, the only constraint that changes is~\eqref{eq:branches}. Increasingly smaller values of $\Lambda$ puts even more stringent bounds on the $g - m_Q$ parameter space; however, the allowed region is centered on more or less the same values. The effect of decreasing $\Lambda$ is illustrated in Figure~\ref{fig:gmq} (right panel), and we find that no viable parameter space exists for $\Lambda \lesssim 0.55$.

\begin{figure}[t]
    \centering
    \begin{minipage}{0.49\textwidth}
    \centering
        \includegraphics[width=\textwidth]{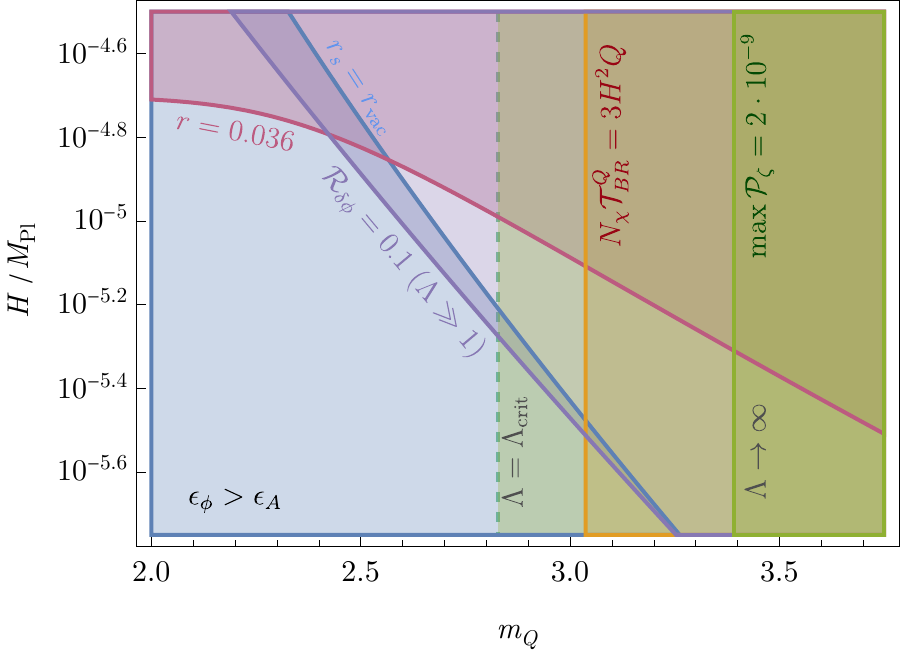}
    \end{minipage}\hfill
    \begin{minipage}{0.49\textwidth}
    \centering
        \includegraphics[width=\textwidth]{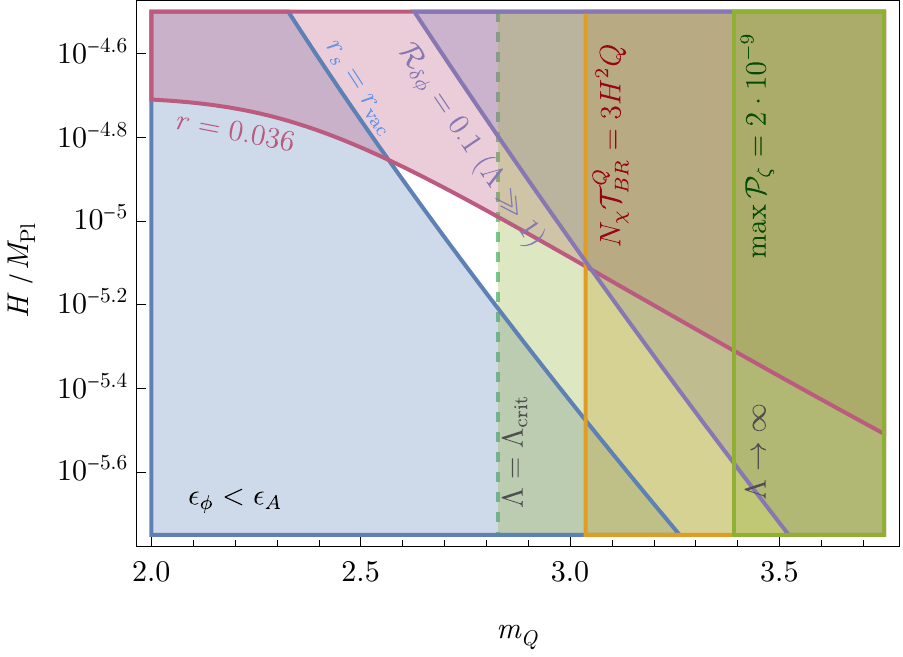}
    \end{minipage}
    \caption{Allowed parameter space of S-CNI in the plane of $m_Q- H/\mpl$.  $\sqrt{2} < m_Q$ is satisfied and $g=0.01$ is fixed  everywhere on both panels. The panels show all constraints discussed in this section in classic (left) and mixed (right) S-CNI scenarios:  the green shaded region with solid border breaks~\eqref{eq:branchesApprox}, the dark yellow region is contaminated by~\eqref{timeclump}, the green shaded region with dashed border violates~\eqref{eq:branches} for $\Lambda \sim \sqrt{2}$, the dark pink region breaks the observational bound  $r<0.036$~\eqref{eq:tensortoscalr}, the chiral signature in the blue region is too weak (see~\eqref{RGG1}), and the purple region is contaminated with too large a non-linear contribution to the scalar power spectrum,~\eqref{eq:noquantum}. At small $\Lambda$ this constraint becomes irrelevant. \textbf{(Left Panel)} For large $\Lambda^2 \gg 1$ values, the $\mathcal{R}_{\delta\phi} >0.1$ constraint kills the available parameter space for enhanced chiral signatures of the \textit{classic} S-CNI branch. \textbf{(Right Panel)} The \textit{mixed} S-CNI branch satisfies all constraints for $6.2 \cdot 10^{-6} \lesssim H/\mpl \lesssim 1.0 \cdot 10^{-5}.$ }\label{fig:branches}
\end{figure}

In Figure~\ref{fig:branches}, we take $g=0.01$ (which lies almost in the center of the narrow interval~\eqref{eq:grange}), and show the overlap of all constraints discussed in this section in the $H - m_Q$ plane. Note that~\eqref{eq:noquantum} is only valid for $\Lambda^2 \gg 1$ and has different expressions depending on which branch of the S-CNI model we pick; it is the only curve that changes between the left (classic S-CNI) and right (mixed S-CNI) panels. It is clear from the Figure that the classic S-CNI branch is unable to satisfy all the constraints at the same time when $\Lambda \gg 1$, but the mixed S-CNI branch does so in a small region of available values for $H/\mpl$:
\begin{align} \label{eq:Hrange}
    6.2 \cdot 10^{-6} \lesssim H/\mpl \lesssim 1.0 \cdot 10^{-5},
\end{align}
and it follows that the same parameter space is available to the classic S-CNI case in the limit of small $\Lambda$ where~\eqref{eq:noquantum} vanishes.

Much like our procedure for Figure~\ref{fig:gmq}, we have drawn a dashed green line to show how~\eqref{eq:branches} would change as one decreases $\Lambda$. The available values for $H/\mpl$ only become stricter for smaller  $\Lambda \sim\sqrt{2}$. This indicates that although~\eqref{eq:noquantum} relaxes in this regime, the parameter space newly opened up in the classic S-CNI scenario may yet be challenged by the more severe~\eqref{eq:branches}.


To summarize, the phenomenological constraints confine three of the parameters in the model to narrow ranges around $g \simeq 0.01$, $H \simeq 8.2 \cdot 10^{-6} \mpl$, and $m_Q \simeq 2.7$. These further imply that $Q \simeq 1.6 \cdot 10^{-3}\mpl$. We also show that for $\Lambda \gg \sqrt{2}$, the only viable branch is the mixed S-CNI scenario. To give some sense of the possible values of other parameters: $\lambda = 60$ and $\Lambda = 10$ would imply $\mu \approx 5.4 \cdot 10^{-4} \mpl$ (which is relatively insensitive to $\Lambda$ provided $\Lambda \gg 1$) and $f \approx 1.4 \cdot 10^{-2} \mpl$.

\section{Assessing the $\Lambda$ Range in S-CNI}
\label{section:Lambda}

From~\eqref{eq:grange},~\eqref{eq:mqrange}, and~\eqref{eq:Hrange}, we have fairly narrow constraints on the parameters $g$, $m_Q$, and $H$. Thus, it is useful to rewrite the equation of motion involving $Q$ in terms of these essentially fixed parameters and $\Lambda$, following definitions in~\eqref{eq:notation}. In this section, we will discuss the allowed values of $\Lambda$, showing that they fall into two regimes, one of which is more promising than the other. The results of this section will play an important role in the constraints that we present in subsequent sections.

Slow-roll solutions have an approximately constant, nonzero, value of $Q$ during the CNI phase, determined by~\eqref{eq:Qdot}. Such a value must be a solution to the polynomial equation
\begin{equation} \label{eq:Qquartic}
    g^2 \lambda^2 Q^4 + 2 g^2 f^2 Q^2 + 2 f^2 H^2 + \frac{g \lambda f}{3 H} U'(\chi) Q = 0,
\end{equation}
where we view $\chi$ as being approximately constant when solving for $Q$. For concreteness, here and in subsequent sections we will take $\chi/f \approx \pi/2$, which maximizes the size of the gravitational wave perturbations (see, e.g., Appendix A of~\cite{Thorne:2017jft-phform}) and sets $U'(\chi) \approx -\mu^4/f$. Viewed as an equation for $Q$,~\eqref{eq:Qquartic} is a messy quartic equation. However, we can rewrite it as a {\em quadratic} equation for $\Lambda$, with $g, \mu^4/f, H,$ and $m_Q$ held fixed:
\begin{equation} \label{eq:Lambdaquadratic}
    \Lambda^2 - \frac{\mu^4/f}{3 H^3 m_Q^2/g} \Lambda + 2 \left(1 + m_Q^{-2}\right) = 0.
\end{equation}
This equation has the two solutions
\begin{equation} \label{eq:Lambdasolutions}
    \Lambda_{1,2} = \frac{\mu^4/f}{6m_Q^2H^3/g}\left[ 1\pm \sqrt{1- 2\ (\mu^4/f)^{-2}(6m_Q^2H^3/g)^{2} \ (1+m_Q^{-2})}\ \right]. 
\end{equation}
Because $\Lambda$ is real by definition, we immediately see that a viable parameter space requires that the argument of the square root is positive, i.e., that
\begin{equation}\label{eq:mu4fmin}
    \frac{\mu^4}{f} \geq \frac{6\sqrt{2} m_Q H^3}{g} \sqrt{m_Q^2 + 1} \equiv \eta H^3,
\end{equation}
where we have defined 
\begin{equation} \label{eq:eta}
    \eta \equiv \frac{6\sqrt{2} m_Q}{g} \sqrt{m_Q^2 + 1} \approx 4.7 \cdot 10^3,
\end{equation}
for benchmark values $g = 0.014$ and $m_Q = 2.7$.
In particular, this gives a lower bound on the larger root $\Lambda_1$, saturated when the square root is zero, and a similar upper bound on the smaller root $\Lambda_2$,    
\begin{align} \label{eq:Lambda1lowerbound}
    \Lambda_1 &\geq \Lambda_{\text{crit}} \equiv \sqrt{2}\ [1+m_Q^{-2}]^{1/2} > \sqrt{2}\ , \\
    \Lambda_2 &\leq \Lambda_{\text{crit}}. \nonumber
\end{align}
The first inequality selects $\Lambda_1$ as the solution to the region of the parameter space where $\Lambda \gg \sqrt{2}$ is satisfied. If we are to explore regions of the parameter space where $\Lambda \leq \sqrt{2}$, we need to pick $\Lambda_2$ as the solution relating $\lambda$ to $f$ and $\mu$. In the S-CNI literature, most of the attention has been directed toward the $\Lambda_1$ solution; as already noted in Section~\ref{sec:previousconstraints}, assuming $\Lambda \gg \sqrt{2}$ simplifies the analytic  expressions.
 
A good solution of~\eqref{eq:Qquartic} should be an attractor of the $Q$ equation of motion, i.e., a value at which $W_\mathrm{eff}''(Q) > 0$ (see~\eqref{eq:Qdot}). Taking a derivative of the right-hand side of~\eqref{eq:Qdot} and substituting the value of $\mu^4/f$ determined by~\eqref{eq:Qquartic}, one finds that a given solution is a minimum provided that
\begin{equation}
    \Lambda^2 \geq \frac{2}{3} \left(m_Q^{-2} - 1\right), 
\end{equation}
a condition that is always satisfied in our parameter space of interest~\eqref{eq:notachyon}. Hence the solutions~\eqref{eq:Lambdasolutions} are always good minima of the effective potential.

The solution $\Lambda_2$ can correspond to small values of $\lambda$, which may not provide sufficiently many e-folds of S-CNI physics~\eqref{eq:lambdarequirement} because $\chi$ rolls too quickly. Another consequence of fast-rolling $\chi$ is a modification of the scalar power spectrum. In fact, combining~\eqref{eq:branches} and~\eqref{eq:branches_Ldep}, we find a constraint 
\begin{equation} \label{eq:Lambdalowerbound}
   \Lambda \geq \Lambda_{\min} \equiv \frac{1 + m_Q^2}{g} 8\pi \sqrt{P_\zeta}\ \bigg[1-\frac{32\pi^2\mathcal{P}_{\zeta}}{g^2}\ m_Q^2(1+m_Q^2)\bigg]^{-1/2} \simeq 0.74 , 
\end{equation}
using benchmark values ($g = 0.014$, $m_Q = 2.7$) and the measured value $P_\zeta^{\rm obs}$ in the last step. Thus we see that, given the requirement of a valid solution with a scalar power spectrum matching observational data, the $\Lambda_2$ regime offers only limited room below $\Lambda = \sqrt{2}$.


Note that Figure~\ref{fig:gmq} (right panel) has made clear that the constraints~\eqref{eq:RGW_Prime},~\eqref{timeclump}, and~\eqref{eq:branches} represent severe restrictions on $g$, $m_Q$, and indeed $\Lambda$. We see that the lowest viable value of $\Lambda$ is $\simeq 0.55$, where the parameter space vanishes to a point; said point is extremal in the $g - m_Q$ plane but the most accomodating in $\Lambda$, allowing for all $\Lambda \geq \Lambda_{\rm min} \simeq 0.55$. Our benchmark values are chosen such that they are comparatively central in the allowed $g - m_Q$ region, but a relatively large range of $\Lambda \lesssim \Lambda_{\rm crit}$ can be accommodated therein.  

This is a significant result which is independent of any UV completion: Even though, unlike the original CNI, $\Lambda \lesssim \sqrt{2}$ was hypothesized as a viable parameter space for S-CNI, not a lot of freedom has been found upon exploration.  

It is worth noting, however, that the blue and orange regions carved out in Figure~\ref{fig:gmq} are not strictly speaking ``no-go'' zones, and thus smaller values of $\Lambda$ are likewise in principle possible to achieve. However, more work must be done in order to predict or discover S-CNI in our universe: If the chiral tensor modes are too small (violating~\eqref{eq:RGW_Prime}), then precise data and analysis techniques are required to extract any CNI  signal. If they are too large (violating~\eqref{timeclump}), then a more robust treatment of these equations must be developed before predictions can be made.

\section{The S-CNI Model and Clockwork}
\label{section:clockwork}

In Section~\ref{sec:previousconstraints}, we detailed constraints on the S-CNI model based primarily on physical consistency with our observed universe. Having cut out much of the na\"ive parameter choices, in this section we examine whether the remaining parameter space admits a UV completion in terms of the clockwork scenario~\cite{Choi:2014rja,Choi:2015fiu,Kaplan:2015fuy}. The analogous question in the original CNI model was studied in~\cite{Reece:Chrono}, which provides the basis for our analysis.

\subsection{Clockwork Constraints}

As discussed in Section~\ref{subsec:period}, the axion periodicity $\chi \cong \chi + 2 \pi f_\chi$ imposes a quantization condition on the coupling: $\lambda = \frac{j\cdot k\cdot g^2}{8\pi^2}$, where $k$ and $j$ are integers generated from different physics in the UV completion. The integer $k$ can arise, for example, from integrating out fermions carrying SU(2) gauge charge that have $\chi$-dependent masses. In order for such an effective theory of fermions and gauge fields to be valid, we require $k < 4\pi/g^2$~\cite{Reece:Chrono}.

The clockwork mechanism~\cite{Choi:2014rja,Choi:2015fiu,Kaplan:2015fuy} is a way to obtain a large integer $j$ as a product of smaller integers. It is, essentially, an iterated version of an older idea of ``axion alignment''~\cite{Kim:2004rp}. One has a set of $n$ axions $\chi_n$ with a potential of the form
\begin{equation} \label{eq:CWpotential}
  \sum_{i=1}^{n-1} \mu_{i+1}^4 \cos\left[\frac{m_i \chi_i}{f_i} + \frac{\chi_{i+1}}{f_{i+1}}\right] + \mu_1^4 \cos\frac{\chi_1}{f_1},
\end{equation}
with $m_i$ a small integer. Integrating out the heaviest modes leads to one parametrically light mode $\chi$ with an effective potential having $j = \prod m_i$. In this way, one can obtain an exponentially large enhancement of $\lambda$ through a product of small integers.

In this setting, a bound can be derived using the fact that each cosine term in~\eqref{eq:CWpotential} mediates axion scattering, and perturbative unitarity provides an upper bound on the corresponding ratios like $\mu_{i+1}^4/f_i^4$. This was shown to imply $j < f/\mu$~\cite{Reece:Chrono}, leading to the conclusion
\begin{equation}\label{eq:clockworklambda}
    \lambda \leq \frac{f}{2\pi \mu}.
\end{equation}
   
\subsection{A Lower Bound on $f$ from Clockwork}

The clockwork constraint~\eqref{eq:clockworklambda} gives an upper bound on $\mu$, for fixed $f$ and $\lambda$. We previously derived a lower bound on $\mu$,~\eqref{eq:mu4fmin}, simply from the existence of a stationary value of $Q$ in the equation of motion. We can combine these two constraints to obtain a range of $\mu$,
\begin{equation} \label{eq:murange}
    (\eta H^3 f)^{1/4} \leq \mu \leq \frac{f}{2\pi \lambda} \lesssim \frac{f}{4 \pi \xi N_\chi},
\end{equation}
where in the last step we used the requirement~\eqref{eq:lambdarequirement} that $\lambda$ be large enough to allow for $N_\chi$ e-folds of $\chi$ rolling.
Requiring that the range~\eqref{eq:murange} is not empty now translates into a {\em lower bound} on $f$,
\begin{equation} \label{eq:flowerboundCW}
    f \gtrsim \eta^{1/3} \left(4\pi \xi N_\chi\right)^{4/3} H \simeq 4.7 \cdot 10^4 H,
\end{equation}
where we substituted $m_Q = 2.7$, $g = 0.014$, and $N_\chi = 10$ in the last step for a numerical estimate.
Together with the range~\eqref{eq:Hrange} of allowed Hubble values, this already shows that achieving a viable clockwork scenario would require $f$ to be very near the Planck scale.
   
\subsection{Restricting the $\Lambda$ Range with Clockwork} 

We can also rearrange our existing constraints to obtain an {\em upper} bound on $f$, assuming a lower bound on $\Lambda$. In the $\Lambda_1$ regime, we have such a bound from the basic constraint~\eqref{eq:Lambda1lowerbound}. In the $\Lambda_2$ regime, we have the lower bound~\eqref{eq:Lambdalowerbound} that originated from the power spectrum constraint~\eqref{eq:branches}. By rewriting the clockwork constraint~\eqref{eq:clockworklambda} in terms of $\Lambda$, we can express this as a range of allowed $\Lambda$ values:
\begin{equation}\label{eq:CWLambda}
    \Lambda_\mathrm{min} \leq \Lambda \leq \Lambda_{\max}^\mathrm{CW}(\mu),
\end{equation}
where 
\begin{equation}
    \Lambda_{\max}^\mathrm{CW}(\mu) = \frac{m_Q H}{2\pi g \mu}.
\end{equation}
Imposing $\Lambda_\mathrm{min} \leq \Lambda_{\max}^\mathrm{CW}(\mu)$ gives an upper bound on $\mu$, which we can combine with the lower bound~\eqref{eq:mu4fmin} $\mu^4 \geq \eta H^3 f$ to obtain
\begin{equation}
(\eta H^3 f)^{1/4} \leq \mu \leq \frac{m_Q H}{2\pi g \Lambda_\mathrm{min}}.   
\end{equation}
Because we have eliminated the $f$-dependence on the right-hand side by rewriting $\lambda/f$ in terms of $\Lambda$, we thus learn that the existence of a clockworkable range of $\Lambda$ implies an {\em upper bound} on $f$:
\begin{equation} \label{eq:fupperboundCW}
    f \leq H \frac{m_Q^2}{6\sqrt{2} (2\pi)^4 g^3 (1 + m_Q^{-2})^{1/2}\Lambda_\mathrm{min}^4} \simeq 620 H,
\end{equation}
again using $m_Q = 2.7$ and $g = 0.014$ in the last step for a numerical estimate and taking $\Lambda_\mathrm{min} \simeq 0.74$ from~\eqref{eq:Lambdalowerbound}. In the $\Lambda_1$ branch, a stronger constraint (by about a factor of 17) can be obtained using $\Lambda_\crit$ for $\Lambda_\mathrm{min}$.

The upper bound~\eqref{eq:fupperboundCW} directly clashes with the lower bound~\eqref{eq:flowerboundCW} that we derived above, showing that there is no clockworkable parameter range.
        
\subsection{Numerical Results}

\begin{figure}[t]
    \centering
    \begin{minipage}{0.49\textwidth}
        \centering
        \includegraphics[width=\textwidth]{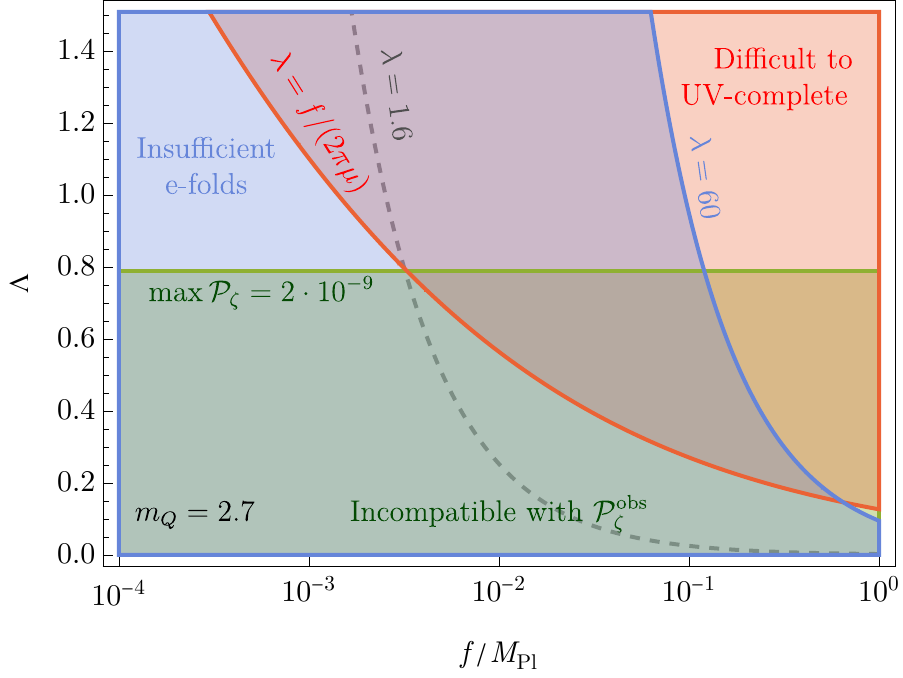}
    \end{minipage}\hfill
    \begin{minipage}{0.49\textwidth}
    \centering                
        \includegraphics[width=\textwidth]{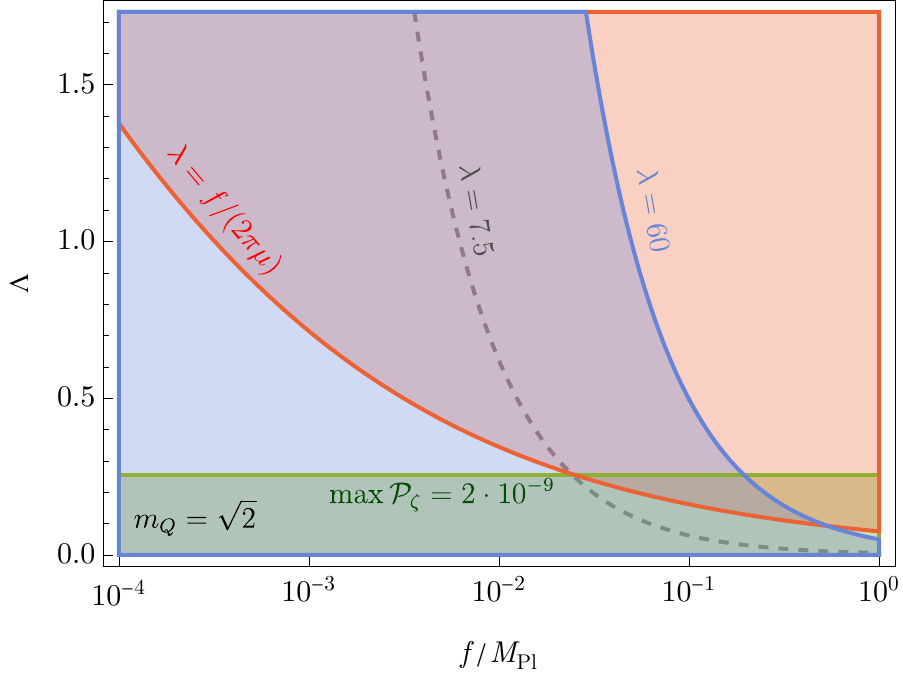}
    \end{minipage}
    \caption{Exploring the $\Lambda_2 \leq \Lambda_{\crit}\sim \sqrt{2}$ (see~\eqref{eq:Lambda1lowerbound}) solutions with various constraints. $g= 0.014$ and $H= 8.2 \cdot 10^{-6}$ for both panels. The red shaded region is forbidden by clockwork~\eqref{eq:Lambda2CW}. The blue shaded region excludes point with $N_{\chi}<10$~\eqref{eq:lambda_const_contour}. The green shaded area doesn't produce correct values for $\mathcal{P}_{\zeta}$. The small area at the bottom right of both panels, trapped between the red and blue curves, corresponds to $f$ values from~\eqref{eq:flowerboundCW}. Not that all regions outside of the blue shaded area correspond to $f \sim 10^{-1} \mpl$. \textbf{(Left Panel)} Choosing the benchmark value $m_Q =  2.7$ allows us to lower $\Lambda_{\min}$ down to $0.74$ (see~\eqref{eq:Lambdalowerbound}). No clockworkable solution with $\lambda \geq 60$ and correct $\mathcal{P}_{\zeta}$ exists. \textbf{(Right Panel)} Lowering benchmark values to $m_Q = \sqrt{2}$ doesn't resolve the UV completion challenge.} \label{fig:LambdaS2}
\end{figure}

To better understand how the clockwork upper bound on $f$ clashes with the lower bound on $\Lambda$, we present a graphical rendition of our equations here. In Figure~\ref{fig:LambdaS2} (left panel), the red shaded region is prohibited by the clockwork constraint. We can find the inequality that describes this region by eliminating $\mu$ from the right hand side of~\eqref{eq:CWLambda}. We replace $\mu$ in terms of $f$ and $\Lambda$ by solving~\eqref{eq:Lambdaquadratic} and find
\begin{align}\label{eq:Lambda2CW}
   f \leq \frac{H m_Q^2/g^3}{48\pi^4\Lambda^3 \big[\Lambda^2+2(1+m_Q^{-2})\big]}, 
\end{align}
as the expression for the clockwork region in the $f -\Lambda$ plane. Notice that this is a $\Lambda$-dependent upper bound on $f$, which becomes arbitrarily weak at small $\Lambda$; this is in contrast to~\eqref{eq:fupperboundCW} above, which used the minimum $\Lambda$ from the power spectrum to give a $\Lambda$-{\em independent} bound on $f$. If we meet the condition~\eqref{eq:Lambda2CW} while staying above contours of $\lambda =60$, we have found the region which satisfies the lower bound~\eqref{eq:flowerboundCW}. The $\lambda=\text{const.~}$contours linearly decay as $f$ gets closer to $\mpl$. We exclude contours with smaller values for $\lambda$ in our plot by
\begin{align}\label{eq:lambda_const_contour}
    \Lambda f = \lambda \ m_QH/g\ \gtrsim 60 \cdot  m_QH/g.
\end{align}
The only region meeting all these requirements resides at the very bottom right of the left panel. The $f$ values residing in this region are extremely close to $\mpl$, and, indeed, coincide with the same ones that satisfy our clockwork lower bound on $f$. 

Both panels in Figure~\ref{fig:LambdaS2} span the range $0\leq \Lambda < \Lambda_{\crit}$. Consequently, $\Lambda_1 \geq \Lambda_{\crit}$ solutions have no chance of being clockworked. 

Now, remember the green shaded region is forbidden for our $\Lambda_2$ solutions because it fails to produce the correct $\mathcal{P}_{\zeta}$ value; hence, $\Lambda_2$ solutions are also kept far from the bottom right region. Notice that even if we had picked different $g$ and $m_Q$ values from~\eqref{eq:grange} and~\eqref{eq:mqrange} to minimize the solid green border to $\Lambda_{\min} \simeq 0.55$ (see Figure~\ref{fig:gmq} right), we would still be too far above the desired values. 

It's worth noting that even without clockwork, taking $\Lambda \lesssim \sqrt{2}$ puts $f$ within an order of magnitude of $\mpl$. This only happens if one abides by $\lambda \gtrsim 60$ and the observed value of $\mathcal{P}_{\zeta}$. So, in addition to accounting for a large coefficient of the Chern-Simons term, justifying this near Planck scale value of $f$ is a second concern that alternative UV completions in this regime have to address.

To provide yet another point of view, observe that for our choice of benchmark values, the gray $\lambda$ contour indicates S-CNI doesn't enter the clockworkable regime until $\lambda \lesssim 1.6$. This value for $\lambda$ doesn't make for enough $N_{\chi}$ to observe any interesting phenomenology. We have also checked that numerical solutions of the classical equations of motion~\eqref{xieom},~\eqref{Qeom},~\eqref{eq:slowrollsum}, and~\eqref{eq:phieom} in this parameter regime are inconsistent with slow-roll evolution of $\chi$ and $Q$. 

Breaking the $g$ upper bound from~\eqref{eq:grange} results in losing control over the backreaction effects (see Figure~\ref{fig:gmq} left), but one may break the $m_Q$ lower bound~\eqref{eq:mqrange} with the aim of obtaining even smaller values for $\Lambda_{\min}$. Taking $m_Q \lesssim 2.7$ violates the $\mathcal{R}_{GW}>1$ bound and results in a less chiral signal. We lowered the bound to $m_Q = \sqrt{2}$ (see Figure~\ref{fig:LambdaS2} right) below which we face tachyonic instabilities in scalar perturbations. Even at this new minimum, S-CNI doesn't become clockworkable until $\lambda \lesssim 7.5$ (gray dashed contour). As mentioned earlier, this $\lambda$ is still too small to account for enough e-folds of interesting phenomenology. 

To summarize, even venturing beyond the $\Lambda >\sqrt{2}$ limit and localizing to small $\Lambda_2$ solutions still results in a large Chern-Simons coefficient that cannot be clockworked. 
We conclude that  S-CNI is not compatible with a clockwork UV completion in any region of its available parameter space.

\section{Additional Constraints}
\label{section:additional}

In this section, we put clockwork aside and consider whether the remaining parameter space from \ref{sec:lowEFTsumm} is well-behaved as a quantum field theory, and challenge this phenomenologically-viable region with more fundamental (albeit sometimes conjectural) criteria. First, we require the couplings and scales that describe the effective theory of S-CNI to respect partial wave unitarity bounds. Next, we consider what portions of the remaining parameter space are consistent with the weak gravity conjecture. 

\subsection{Perturbative Unitarity Bounds}
        
The partial wave unitarity bound states that EFT amplitudes cannot be arbitrarily large, if we want our perturbative approach to remain reliable. The perturbative unitarity bound on the $\chi \chi \to \chi \chi$ scattering amplitude arising from the cosine potential for $\chi$ was previously discussed in~\cite{Reece:Chrono}, and played a role in the derivation of the bound~\eqref{eq:clockworklambda} on $\lambda$ in the clockwork scenario. In this section we will give a more {\em direct} bound on $\lambda$, based on the scattering of gauge fields and axions mediated by the $\lambda \chi F \widetilde{F}$ coupling. Similar perturbative unitarity bounds on such a coupling have recently been derived (with very different applications in mind) in Refs.~\cite{Brivio:2021fog, Inan:2022rcr}. The basic idea of such bounds is to use the unitarity of the $S$-matrix, together with a decomposition of scattering states into angular momentum modes, to bound partial-wave amplitudes which are functions of center-of-mass energy $\sqrt{s}$ alone. Specifically, for a $2 \to 2$ scattering process of states with momentum along the $z$ axis and incoming helicities $\lambda_1, \lambda_2$ into outgoing states at angles $(\theta, \phi)$ with outgoing helicities $-\lambda_3, -\lambda_4$, the amplitude decomposes as
\begin{align}
    {\cal M}_{\{\lambda_i\}}(s,\theta,\phi)= 16\pi \sum_j (2j+1) \sqrt{(1+\delta_{\lambda_1 \lambda_2})(1+\delta_{\lambda_3\lambda_4})} e^{i (\lambda + \mu)\phi} d^j_{\lambda,-\mu}(\theta) {\cal T}^j_{\{\lambda_i\}}(s). 
\end{align}
Here $\lambda = \lambda_1 - \lambda_2$ and $\mu = -(\lambda_3 - \lambda_4)$. The combination $e^{i (\lambda + \mu)\phi} d^j_{\lambda,-\mu}(\theta)$ corresponds to a Wigner $D$-function evaluated on this specific kinematic configuration. The $D$-functions have simple orthogonality properties, so it is possible to read off a given partial-wave amplitude ${\cal T}^j_{\{\lambda_i\}}(s)$ by integrating the full amplitude $\cal M$ against a $D$-function. In the special case $\lambda = \mu = 0$, these reduce to the familiar Legendre polynomials. The partial wave unitarity bound constrains the size of ${\cal T}^j_{\{\lambda_i\}}(s)$ for an {\em elastic} scattering process, at any $s$:
\begin{equation}
    |{\cal T}^j(s) - i/2| \lesssim 1/2.
\end{equation}
In particular, for the amplitudes we will consider where the leading tree-level contribution to ${\cal T}^j(s)$ is real, we will demand that this contribution be smaller than $1/2$. For more details, a textbook discussion appears in~\cite{Itzykson:1980rh}.
        
To generalize the partial wave unitarity bound for the S-CNI model, we computed the tree level scattering amplitudes of gauge field axion scattering from
\begin{equation}
    \scalebox{0.75}{
    \begin{tikzpicture}[baseline=(a.base)]
        \begin{feynman}
            \vertex (i1){$\epsilon_1$};
            \vertex [below=2cm of i1] (i2){$p_2$};
            \vertex [right=1cm of i1] (O1);
            \vertex [below=1cm of O1] (a);
            \vertex [right=1cm of a] (b);
            \vertex [right=3cm of i1] (f3){$\epsilon_3$};
            \vertex [below=2cm of f3] (f4){$p_4$};
            \diagram*{
                (i1) -- [photon,momentum={[arrow shorten=0.35] \(\)} ] (a),
                (i2) -- [scalar,momentum'={[arrow shorten=0.35] \(\)}] (a),
                (a) -- [photon] (b),      
                (f3) -- [photon, momentum'={[arrow shorten=0.35] \(\)}] (b),
                (f4) -- [scalar,momentum={[arrow shorten=0.35] \(\)}] (b),
                };
        \end{feynman}
    \end{tikzpicture}} \qquad\text{, and} \qquad \scalebox{0.75}{
    \begin{tikzpicture}[baseline=(c.base)]
        \begin{feynman}
            \vertex (i1){$\epsilon_1$};
            \vertex [below=3cm of i1] (i2){$p_2$};
            \vertex [right=1cm of i1] (O1);
            \vertex [below=1cm of O1] (a);
            \vertex [below=0.5cm of a] (c);
            \vertex [below=1cm of a] (b);
            \vertex [right=2cm of i1] (f3){$\epsilon_3$};
            \vertex [below=3cm of f3] (f4){$p_4$};
            \diagram*{
                (i1) -- [photon, momentum'={[arrow shorten=0.35] \(\)}] (a),
                (i2) -- [scalar, momentum={[arrow shorten=0.35] \(\)}] (b),
                (a) -- [photon] (b),
                (f3) -- [photon, momentum={[arrow shorten=0.4] \(\)}] (b),
                (f4) -- [scalar, momentum'={[arrow shorten=0.4] \(\)}](a),
                    };
            \end{feynman}
    \end{tikzpicture}}.
\end{equation}
We have computed these amplitudes using the spinor helicity formalism; in this case, we chose the external $\chi$ to be massless to facilitate the computation (expecting that perturbative unitarity will break down only at energies $\sqrt{s} \gg m_\chi$). We find:
\begin{align}
    \mathcal{M}_{+-}^{A\chi}&= \delta^{ab} \frac{u+s}{su}\left(\frac{\lambda}{2f}\right)^2\VEV{14}^2[34]^2 \propto \delta^{ac} \left(\frac{\lambda}{4f}\right)^2 2s(1-\cos{\theta}),\\
    \mathcal{M}^{A\chi}_{++} &= -2 \delta^{ac} \left(\frac{\lambda}{2f}\right)^2 \VEV{13}^2 \propto \delta^{ab} \left(\frac{\lambda}{4f}\right)^2 4s(1-\cos{\theta}).
\end{align}
These calculations are presented in the all-incoming convention for momenta and helicities, i.e., the subscripts $+$ and $-$ correspond to the incoming helicity of the two gauge fields (particles 1 and 3). The $\delta^{ac}$ factor ensures that the color of the outgoing gauge field is the same as that of the incoming gauge field. We use $\propto$ here to indicate that we have dropped {\em phases}, that is, we have carried out replacements like $\langle 1 4 \rangle \to \sqrt{u}$. Constant prefactors have not been dropped. The angular dependence of the amplitude $\mathcal{M}^{A\chi}_{++}$ is exactly that of the Wigner $D$-function $D^1_{1,-1}$, whereas the amplitude $\mathcal{M}^{A\chi}_{+-}$ has nontrivial overlap with a large set of $D$-functions $D^j_{1,1}$.
        
A second unitarity constraint shows up when considering the gauge boson scatterings that are mediated by the axion,
\begin{equation}
    \scalebox{0.75}{
    \begin{tikzpicture}[baseline=(a.base)]
        \begin{feynman}
            \vertex (i1){$\epsilon_1$};
            \vertex [below=2cm of i1] (i2){$\epsilon_2$};
            \vertex [right=1cm of i1] (O1);
            \vertex [below=1cm of O1] (a);
            \vertex [right=1cm of a] (b);
            \vertex [right=3cm of i1] (f3){$\epsilon_3$};
            \vertex [below=2cm of f3] (f4){$\epsilon_4$};
            \diagram*{
                (i1) -- [photon,momentum={[arrow shorten=0.35] \(\)} ] (a),
                (i2) -- [photon,momentum'={[arrow shorten=0.35] \(\)}] (a),
                (a) -- [scalar] (b),      
                (f3) -- [photon, momentum'={[arrow shorten=0.35] \(\)}] (b),
                (f4) -- [photon,momentum={[arrow shorten=0.35] \(\)}] (b),
            };
        \end{feynman}
    \end{tikzpicture}} \qquad, \scalebox{0.75}{
    \begin{tikzpicture}[baseline=(c.base)]
        \begin{feynman}
            \vertex (i1){$\epsilon_1$};
            \vertex [right=2cm of i1] (f3){$\epsilon_3$};
            \vertex [right=1cm of i1] (O1);
            \vertex [below=1cm of O1] (a);
            \vertex [below=0.5cm of a] (c);
            \vertex [below=1cm of a] (b);
            \vertex [below=3cm of i1] (i2){$\epsilon_2$};
            \vertex [right=2cm of i2] (f4){$\epsilon_4$};
            \diagram*{
                (i1) -- [photon,momentum'={[arrow shorten=0.35] \(\)} ] (a),
                (i2) -- [photon,momentum={[arrow shorten=0.35] \(\)}] (b),
                (a) -- [scalar] (b),      
                (f3) -- [photon, momentum={[arrow shorten=0.35] \(\)}] (a),
                (f4) -- [photon,momentum'={[arrow shorten=0.35] \(\)}] (b),
            };
        \end{feynman}
    \end{tikzpicture}} \qquad\text{, and }\qquad \scalebox{0.75}{
    \begin{tikzpicture}[baseline=(c.base)]
        \begin{feynman}
            \vertex (i1){$\epsilon_1$};
            \vertex [below=3cm of i1] (i2){$p_2$};
            \vertex [right=1cm of i1] (O1);
            \vertex [below=1cm of O1] (a);
            \vertex [below=0.5cm of a] (c);
            \vertex [below=1cm of a] (b);
            \vertex [right=2cm of i1] (f3){$\epsilon_3$};
            \vertex [below=3cm of f3] (f4){$p_4$};
            \diagram*{
                (i1) -- [photon, momentum'={[arrow shorten=0.35] \(\)}] (a),
                (i2) -- [photon, momentum={[arrow shorten=0.35] \(\)}] (b),
                (a) -- [scalar] (b),
                (f3) -- [photon, momentum={[arrow shorten=0.4] \(\)}] (b),
                (f4) -- [photon, momentum'={[arrow shorten=0.4] \(\)}](a),
            };
        \end{feynman}
    \end{tikzpicture}}.
\end{equation}
In this case we use spinor helicity for the external gauge bosons while keeping the mass in the internal propagator, finding: 
\begin{align}
    \mathcal{M}_{++--}^{AA}&=\left(\frac{\lambda}{2f}\right)^2 \frac{\delta^{ab}\delta^{cd}}{s-m_{\chi}^2}\ \VEV{12}^2[34]^2\propto \left(\frac{\lambda}{4f}\right)^2 \frac{4s^2 \delta^{ab}\delta^{cd}}{s-m_{\chi}^2},\\
    \mathcal{M}_{+-+-}^{AA}&=\left(\frac{\lambda}{2f}\right)^2 \frac{\delta^{ac}\delta^{bd}}{t-m_{\chi}^2}\ \VEV{24}^2[13]^2\propto \left(\frac{\lambda}{4f}\right)^2 \frac{4t^2 \delta^{ac}\delta^{bd}}{t-m_{\chi}^2} \approx -\left(\frac{\lambda}{4f}\right)^2 \delta^{ac}\delta^{bd} 2s(1-\cos{\theta}),\\
    \mathcal{M}_{+--+}^{AA} &= \left(\frac{\lambda}{2f}\right)^2 \frac{\delta^{ad}\delta^{bc}}{u-m_{\chi}^2}\ \VEV{23}^2[14]^2\propto \left(\frac{\lambda}{4f}\right)^2 \frac{4u^2 \delta^{ad}\delta^{bc}}{u-m_{\chi}^2} \approx -\left(\frac{\lambda}{4f}\right)^2 \delta^{ad}\delta^{bc} 2s(1+\cos{\theta}),\\
    \mathcal{M}_{++++}^{AA} &= -\left(\frac{\lambda}{4f}\right)^2 \left[ \frac{\delta^{ab}\delta^{cd}}{s-m_{\chi}^2}[12]^2[34]^2 + \frac{\delta^{ac}\delta^{bd}}{t-m_{\chi}^2}[34]^2[13]^2 + \frac{\delta^{ad}\delta^{bc}}{u-m_{\chi}^2}[14]^2[23]^2 \right].
\end{align}
Again, we have used the all-incoming convention for momenta and helicities; $a$, $b$, $c$, and $d$ are the colors of the gauge fields $1$, $2$, $3$, and $4$. Expressions following $\propto$ have dropped phases, while expressions following $\approx$ have been taken to the high-energy limit and hence set $m_\chi \to 0$.
        
Simply by dimensional analysis, all of these amplitudes grow with energy roughly as $\lambda^2 E^2 / f^2$. However, the precise bound on a partial wave amplitude depends on the constant coefficient as well as the overlap of the angular structure in $\cal M$ with the appropriate Wigner $D$-function for the partial wave. From inspecting the leading growth with $s$ of the elastic scattering processes we have computed, we find that the amplitude $\mathcal{M}^{AA}_{++--}$ with all four gauge fields of the same color leads to a relatively stringent bound:
\begin{equation} \label{eq:unitarity}
    \sqrt{s} \leq 8 \sqrt{\pi} \frac{f}{\lambda}.
\end{equation}
This is a bound on the center-of-mass energy at which the $2 \to 2$ scattering process can be approximately described by the EFT. If we further suppose that the physics describing the origin of the scale $\mu$ should be describable within an EFT containing the $\lambda \chi F \widetilde{F}$ term, we would conclude that
\begin{equation} \label{eq:muunitarity}
    \mu \lesssim 8 \sqrt{\pi} \frac{f}{\lambda}.
\end{equation}
It is intriguing that this bound has the same parametric form as the constraint~\eqref{eq:clockworklambda} arising in clockwork models. However, it is quantitatively weaker; the coefficient $8\sqrt{\pi}$ is larger than the coefficient $1/(2\pi)$ by about two orders of magnitude. By scattering more general states which are superpositions of different colors and helicities and finding eigenvalues of the ${\cal T}$ matrix, it is possible that we could refine the perturbative unitarity bound by an $O(1)$ factor. However, we do not expect that a bound along these lines can be nearly as strong as the clockwork bound. The prefactor in the bound on $\mu$ appeared to the $4/3$ power in the lower bound on $f$~\eqref{eq:flowerboundCW} and to the {\em fourth} power in the upper bound on $f$~\eqref{eq:fupperboundCW}. Thus, unlike the clash between these bounds that arose in the clockwork-based argument, there are many orders of magnitude of $f$, from $\sim 100H$ to $\sim 10^{10} H$, allowed by the perturbative unitarity bound~\eqref{eq:muunitarity}.

Although the perturbative unitarity constraint on the scattering of axions and gauge bosons is not strong enough by itself to eliminate the interesting S-CNI parameter space, we present the bounds here in case they could form a useful starting point for further exploration. In recent years, the general logic of unitarity bounds has been extended in the direction of quite powerful $S$-matrix bootstrap and positivity bound techniques, which might be interesting to apply to the S-CNI model in the future (see, e.g.,~\cite{Kruczenski:2022lot,deRham:2022hpx} for recent surveys).

\subsection{The Weak Gravity Conjecture Constraint}
    
Theories of quantum gravity are expected to lack any global symmetries (see, e.g.,~\cite{Banks:2010zn, Harlow:2018tng}). From this general idea, a number of quantitative (but conjectural) constraints on EFTs coupled to gravity have emerged in recent years. These potentially lead to important constraints on variations of chromo-natural inflation, though as we will see, obtaining strong constraints requires going beyond the most minimal conjectures. 
    
The Weak Gravity Conjecture (WGC) provides a quantitative explanation of what goes wrong in a theory that has an approximate global symmetry due to a small gauge coupling. In particular, the magnetic WGC holds that an effective gauge theory with coupling $g$ must break down at energies $\Lambda_\mathrm{UV} \lesssim g M_\mathrm{Pl}$~\cite{Arkani-Hamed:2006emk}. Although the original form of this conjecture applied only to U(1) gauge theories, subsequent work has argued that it should apply to nonabelian gauge theories as well~\cite{Heidenreich:2015nta,Heidenreich:2017sim} (see also~\cite{Cota:2020zse} and the review~\cite{Harlow:2022gzl}). Because the original (non-spectator) CNI model required an extremely small gauge coupling, $g \sim 10^{-6}$, it ran into immediate tension with the WGC~\cite{Heidenreich:2017sim}.
    
The spectator CNI models, preferring $g \sim 10^{-2}$~\eqref{eq:grange}, are somewhat less constrained, but we still conclude from this that the EFT should break down at or below the scale
\begin{equation}
    \Lambda_\mathrm{UV} \lesssim g M_\mathrm{Pl} \lesssim 4 \cdot 10^{16}\,\mathrm{GeV}.
\end{equation}
While the minimal WGC arguments do not specify precisely what happens at this scale, in practice it is generally a radical departure from 4d local EFT (e.g., extra dimensions or high-spin string states appear), so we should require 4d mass scales like $\mu$ and $H$ to be below this bound. The parameter space of interest satisfies this constraint with room to spare.
    
The WGC also constrains theories of axions. The minimal axion WGC requires that, for an axion $\chi$ of decay constant $f_\chi$, there must be instantons with action $S_\chi \lesssim n_\mathrm{inst} M_\mathrm{Pl}/f_\chi$. Checking this constraint on our model is not completely straightforward, without knowing the microscopic origin of the $\mu^4$ factor in the axion potential. However, it is reasonable to expect that $\mu^4 \sim \Lambda_\mathrm{UV}^4 \exp(-S_\chi / b)$ where $S_\chi$ is an instanton action, for some $O(1)$ coefficient $b$, and thus $S_\chi \sim O(1) \log(\Lambda_\mathrm{UV}/\mu)$. In the parameter space of interest (see the end of Section~\ref{sec:lowEFTsumm}), $\log(M_\mathrm{Pl}/\mu) \sim 7$, which is smaller than $M_\mathrm{Pl}/f \sim 10^{2}$. Thus, the axion WGC bound is expected to be satisfied by whatever UV physics generates the axion potential.\footnote{We have $1/f_\chi = j/f$; naively the cosine potential appears to have $n_\mathrm{inst} = 1/j$, but this can arise from fundamental instantons of charge $(0,1)$ and $(1,j)$ in a two-axion model~\cite{Kim:2004rp}. These considerations do not seem to alter the conclusion, at least without a detailed model in which the $O(1)$ coefficients in the bound can be checked.} There is also a {\em magnetic} axion WGC, which requires the existence of an axion string (around which the $\chi$ field winds) with tension $T_\chi \lesssim f_\chi \mpl = f \mpl/j$. Perturbativity of the gauge theory requires that $k g^2/(4\pi) \lesssim 1$, which requires that $j \gtrsim 2\pi \lambda \gtrsim 4 \cdot 10^2$. If $f \sim 10^{-2} \mpl$, this implies that the string mass scale is $\sqrt{2\pi T_\chi} \lesssim 10^{-2} \mpl$. This is safely larger than $\mu$ or $H$, so there is no reason to expect the axion string modes to alter the dynamics of the model. Thus, we conclude that the S-CNI model can survive both the electric and magnetic forms of the axion WGC.
    
A potentially more interesting constraint arises from the conjecture that quantum gravity requires axions to eliminate a $(-1)$-form Chern-Weil symmetry associated with the SU(2) instanton number~\cite{Heidenreich:2020pkc}. The axion $\chi$ couples to two different kinds of instantons: those responsible for generating the $U(\chi)$ term, and the SU(2) instantons. Unless these two instantons are related in the UV, e.g., through unification (unlikely, since they have very different actions), this means that $\chi$ gauges only one linear combination of the two $(-1)$-form symmetries. Another axion $\theta$, then, would be required to eliminate the remaining SU(2) instanton number symmetry. The axion WGC applied to this other axion would imply that it has a decay constant $f_\theta \lesssim \frac{g^2}{8\pi^2}M_\mathrm{Pl} \sim 3 \cdot 10^{12}\,\mathrm{GeV} \left(\frac{g}{10^{-2}}\right)^2$. This is a more interesting constraint, because it is close to the scale $H/(2\pi)$ inferred from~\eqref{eq:Hrange}. Hence, this hypothetical second axion would fluctuate significantly during inflation, leading to the production of strings around which $\theta$ winds; call these $\theta$-strings. The magnetic axion WGC for $\theta$ indicates that these would have tension
\begin{equation} \label{eq:TthetaboundWGC}
T_\theta \lesssim f_\theta \mpl \lesssim \frac{g^2}{8\pi^2} \mpl^2.
\end{equation}
The production of loops of $\theta$-string has an energy cost, which should not form a significant source of backreaction on inflation. We can estimate this as follows: in a region of size $L$, we expect the typical fluctuation of the field  $\theta$ over one Hubble time to be of order $\frac{H}{2\pi} (HL)$. We form a string when this is of order $f_\theta$, which occurs in a region of length $L_* \sim \frac{2\pi f_\theta}{H^2}$, with energy $T_\theta L_*$. This should be compared to the rate at which energy is ordinarily depleted from a region of volume $L_*^3$ during one Hubble time, $\sim \epsilon_H (H^2 M_\mathrm{Pl}^2) \frac{L_*^3}{H}$. Demanding that the string cost is small compared to the ordinary depletion of energy leads to the bound
\begin{equation} \label{eq:Tthetabound}
    T_\theta \ll \epsilon_H \frac{(2\pi f_\theta)^2}{H^2} M_\mathrm{Pl}^2.
\end{equation}
We expect the $\theta$-string mass scale $M_\theta \equiv \sqrt{2\pi T_\theta}$ to be a strong UV cutoff, in the sense that local QFT breaks down at this scale due to the appearance of fundamental high-spin states (see~\cite{Heidenreich:2021yda}). In the preferred parameter space of the S-CNI model, $\epsilon_H \sim 10^{-5}$, and the bounds~\eqref{eq:TthetaboundWGC} and~\eqref{eq:Tthetabound} are numerically similar, both implying $M_\theta \lesssim 10^{16}\,\mathrm{GeV}$. Thus, this constraint is not so different from the $g M_\mathrm{Pl}$ WGC bound. However, it does have potentially important consequences: if the theory approximately saturates these bounds, then $\theta$-strings potentially play a role in the inflationary dynamics. These strings would likely persist after inflation, because the SU(2) confinement scale is far below our current Hubble scale, so confinement does not occur to generate string-destroying domain walls. Such strings could be detected through their gravitational signatures; they would also emit SU(2) dark radiation, potentially detectable through $N_\mathrm{eff}$ measurements. (Other sources of SU(2) dark radiation are discussed in~\cite{Kakizaki:2021mgj}.)
    
Since the additional axion $\theta$ relies on a substantially stronger conjecture than the original WGC, we will not pursue a more detailed exploration of its consequences here. The implications for chromo-natural inflation of a completely different set of conjectural Swampland constraints are discussed in~\cite{Montero:2022jrc}.
        
\section{Conclusion}
\label{sec:conclusion}

The chromo-natural inflation model and its relatives offer an appealing scenario that generates a distinctive chiral primordial gravitational wave signal. It is important to understand whether such models can actually be realized as consistent quantum field theories, and if so, whether these can be embedded in quantum gravity. We have built on earlier results~\cite{Reece:Chrono} to argue that there is a substantial challenge to realizing a UV completion of the Spectator CNI model of Ref.~\cite{Dimastrogiovanni:2016fuu}. The basic problem is that the effective coupling $\lambda$ of the axion to SU(2) gauge fields must be large (see Eq.~\eqref{eq:lambdarequirement}), in order to support several e-folds in which the CNI phenomenology is operative. On the other hand, this coupling must be an integer multiple of a perturbatively small loop factor (see~\eqref{eq:CSterminteger}). One attempt to realize the necessary large integer coefficient is for it to arise as a product of smaller integers, through the clockwork mechanism~\cite{Choi:2014rja,Choi:2015fiu,Kaplan:2015fuy}. We have shown, in Section~\ref{section:clockwork}, that this mechanism cannot generate a sufficiently large coupling to satisfy all of the phenomenological and consistency requirements on the spectator CNI model: one can derive two bounds on the axion scale $f$,~\eqref{eq:flowerboundCW} and~\eqref{eq:fupperboundCW}, that are mutually inconsistent by two orders of magnitude.

All of the details of the clockwork scenario, in the end, are subsumed in the statement that $\lambda/f \lesssim \mu/(2\pi)$. This excludes the parameter space of the model. It is intriguing that bounds on $\lambda/f$ can also be deduced purely within effective field theory: as we argued in Section~\ref{section:additional}, perturbative unitarity alone implies that $\lambda/f \lesssim 4\sqrt{2\pi}\mu$. This has the same form as the clockwork constraint, but is numerically weaker, and is insufficient to fully exclude the parameter space of the model. We have also briefly discussed possible constraints from variations on the Weak Gravity Conjecture. As is often the case in such applications, there are potentially interesting bounds but they require stronger conjectures beyond the minimal and most well-established ones.

In light of our results, there are some further potential avenues to explore. One would be to investigate models with a more general potential $U(\chi)$ rather than a simple cosine potential. Axion monodromy~\cite{Silverstein:2008sg,Kaloper:2008fb} could allow for an effectively non-periodic potential, an idea that has been invoked, but not thoroughly explored, in the CNI context~\cite{Maleknejad:2016dci, Maleknejad:2020yys, Maleknejad:2020pec}. Structurally, these models have a similar challenge in explaining the origin of the large coupling of axions to gauge fields, but the details are different enough that the constraints discussed here should be re-evaluated.

Another direction would be to attempt to find a set of assumptions weaker than a clockwork UV completion, but stronger than perturbative unitarity in the EFT, that is sufficient to rule out the model. For example, one could focus on models in which the $\chi F \widetilde{F}$ coupling is generated by integrating out 4d fermions in various representations of the SU(2) gauge group, and study additional higher-dimension operators arising in the EFT and relative constraints on their coefficients. In recent years, various positivity and bootstrap bounds have also been used to strengthen conclusions beyond those available from partial-wave unitarity bounds. These methods could be applied in this context. A quite different possible origin for the $\chi F \widetilde{F}$ coupling is from a higher-dimensional Chern-Simons term, with a large level, potentially arising from a flux through still other extra dimensions. Given the phenomenological requirements of the model, the scales $H$, $\mu$, and $f$ must be sufficiently close to the Planck scale that there is relatively little room for engineering hierarchies of scales involving multiple extra dimensions. Nonetheless, it could be interesting to assess the possibility more quantitatively.

Using the clockwork scenario as an example, we have highlighted the challenges involved in finding a UV-complete theory that explains the origin of parameters in the S-CNI model. We expect that similar challenges would arise in any attempt at building a UV completion. However, we do not have a sharp no-go theorem. Given the potential importance of a chiral gravitational wave signal, we hope to see further creative ideas for new models and mechanisms that could convincingly embed such a signal in a consistent quantum theory.
    
\section*{Acknowledgments}

HB thanks A.~Bedroya and E.~Sussman for helpful discussions. HB and MR are supported in part by the NASA Grant 80NSSC20K0506 and the DOE Grant DE-SC0013607. W.L.X.~thanks the Mainz Institute of Theoretical Physics of the Cluster of Excellence PRISMA+ (Project ID 39083149) for its hospitality during completion of part of this work, and is supported by the U.S.~Department of Energy under Contract DE-AC02-05CH11231.
    
\appendix

\section{Deriving the $\ephi$ Equation of Motion}\label{app:phieom}

In this appendix, we present the derivation of the equation~\eqref{eq:phieom} for the time dependence of $\ephi$ without reference to the form of $V(\phi)$. First, we present some definitions. We will use the differential number of e-folds, $dN = - H dt$, and the slow-roll parameters, both defined in terms of $H(t)$ (with subscript $H$) and in terms of $V(\phi)$ (with subscript $\phi$) (see, e.g., chapter 18 of~\cite{lyth_liddle_2009}):
\begin{align}
    \label{def:epsH}
    \epsilon_H &\equiv - \frac{\dot H}{H^2} = \frac{d\ln{H}}{dN}, \\
    \label{def:etaH}
    \eta_H &\equiv \epsilon_H - \frac{1}{2} \frac{\dot \epsilon_H}{H\epsilon_H} = \epsilon_H +\frac{1}{2}\frac{d\ln{\epsilon_H}}{dN} \\
    \label{def:epsphi}
    \ephi &\equiv \frac{{\dot \phi}^2}{2 H^2 \mpl^2} \approx \frac{1}{2} \mpl^2 \left(\frac{V'(\phi)}{V(\phi)}\right)^2, \\
    \label{def:etaphi}
    \eta_\phi &\equiv \mpl^2 \frac{V''(\phi)}{V(\phi)}.
\end{align}
In the main text, we defined $\ephi$ in~\eqref{eq:slowroll} (as above) in terms of $\dot \phi$, because this is exactly what appears in the classical equation of motion~\eqref{eq:slowrollsum}. The conventional definition, however, is the final term in~\eqref{def:epsphi} in terms of $V'/V$. These are not equivalent, but they are approximately equal, assuming that $\phi$ satisfies the slow-roll condition~\eqref{eq:secondderivs} {\em and} that $V(\phi)$ dominates the energy density of the universe, so that $V(\phi) \approx 3 H^2 \mpl^2$.

We would like to find an equation for the time derivative ${\dot \epsilon}_\phi$ that does not involve the exact form of the potential $V(\phi)$. We will do this in two steps: first, we will find an expression for ${\dot \epsilon}_\phi$ that depends on $V$ only implicitly through $\eta_\phi$. Then, we will relate the spectral index $n_s - 1$ to $\eta_\phi$, allowing us to finally obtain~\eqref{eq:phieom}.

Taking a time derivative of~\eqref{def:epsphi} and using the time derivative of the slow-roll approximation $3 H {\dot \phi} = - V'(\phi)$, we find that 
\begin{equation} \label{eq:dephidt}
 {\dot \epsilon}_\phi \approx 2 H \epsilon_\phi (2 \epsilon_H - \eta_\phi).
\end{equation}
This involves $V(\phi)$ only through the slow-roll parameter $\eta_\phi$. Our next goal is to re-express this in terms of $n_s$.

Consider the spectral tilt
    \begin{equation} \label{eq:nsminusone}
        n_s-1 = \frac{d\ln{\Delta_s^2}}{d\ln{k}} = \frac{d\ln{\Delta_s^2}}{dN}\cdot \frac{dN}{d\ln{k}}.
    \end{equation}
Using the equation $\Delta_s^2 = \frac{H^2}{8\pi^2\mpl^2} \cdot \frac{\epsilon_{\phi}}{\epsilon_H^2}$ from~\cite{Papageorgiou:2019ecb}, we have
    \begin{equation} 
        \frac{d\ln{\Delta_s^2}}{dN} = 2\frac{d\ln{H}}{dN} + \frac{d\ln{\epsilon_{\phi}}}{dN}-2\frac{d\ln{\epsilon_H}}{dN}. \label{spectrum}
    \end{equation}
From the horizon crossing relation $k = aH$, we have 
\begin{equation} \label{eq:dNdlnk}
d\ln{k} = -dN + d\ln{H} \quad \Rightarrow \quad \frac{dN}{d\ln{k}} = \left(-1+\frac{d\ln{H}}{dN}\right)^{-1} = -(1-\epsilon_H)^{-1}.
\end{equation}
From the definition~\eqref{def:etaH}, we have
\begin{equation}\label{eq:dlneHdN}
    \frac{d\ln{\epsilon_H}}{dN} = 2(\eta_H-\epsilon_H). 
\end{equation}
Finally, we can rewrite~\eqref{eq:dephidt} as
\begin{equation} \label{eq:dlnephidN}
 \frac{d\ln{\epsilon_{\phi}}}{dN} = -2(2\epsilon_H - \eta_\phi).
\end{equation}
Plugging in~\eqref{def:epsH},~\eqref{eq:dlnephidN}, and~\eqref{eq:dlneHdN} into~\eqref{spectrum}, and the result together with~\eqref{eq:dNdlnk} into~\eqref{eq:nsminusone}, we find
\begin{equation} \label{eq:nsminusoneresult}
    n_s -1 \approx (4 \eta_H - 2 \eta_\phi - 2 \epsilon_H) \cdot (1 - \epsilon_H)^{-1} \approx 4 \eta_H - 2 \eta_\phi - 2 \epsilon_H,
\end{equation}
up to higher order terms in the slow-roll expansion and corrections due to subdominant contributions to the energy density of the universe.

Finally, we substitute~\eqref{eq:nsminusoneresult} into~\eqref{eq:dephidt} to obtain
\begin{equation}
    {\dot \epsilon}_\phi \approx H \ephi \left(6 \epsilon_H - 4 \eta_H + n_s - 1\right),
\end{equation}
which we quoted in the main text as~\eqref{eq:phieom}, substituting explicit expressions for $\epsilon_H$ and $\eta_H$ in terms of $H(t)$.
    
\bibliographystyle{JHEP}
\bibliography{bib}

\end{document}